\documentclass[lettersize,journal]{IEEEtran}

\usepackage{cite}
\usepackage[utf8]{inputenc} %
\usepackage[T1]{fontenc}    %
\usepackage{hyperref}       %
\usepackage{url}            %
\usepackage{booktabs}       %
\usepackage{amsfonts}       %
\usepackage{nicefrac}       %
\usepackage{microtype}      %
\usepackage{xcolor}         %

\usepackage{amsmath}

\usepackage{amsthm}

\usepackage{latexsym}
\usepackage{bbding}
\usepackage{algorithm}
\usepackage{algorithmic}
\usepackage{float}
\usepackage{multirow}
\usepackage{booktabs}
\usepackage{graphicx}
\usepackage{subcaption}
\usepackage{array}
\usepackage{booktabs}
\usepackage{soul}
\usepackage{pifont}
\usepackage{diagbox}
\usepackage[linewidth=0.5pt]{mdframed}

\usepackage{tabularx}
\usepackage{booktabs}
\usepackage{multicol}

\usepackage{enumitem}
\usepackage{dsfont}
\usepackage{footnote}

\usepackage{arydshln} %

\allowdisplaybreaks

\newtheorem{definition}{Definition}[section]

\newcommand{\PVM}{\textit{PVMark}}



\usepackage{enumitem}
\setenumerate[1]{itemsep=0pt,partopsep=0pt,parsep=\parskip,topsep=5pt}
\setitemize[1]{itemsep=0pt,partopsep=0pt,parsep=\parskip,topsep=5pt}
\setdescription{itemsep=0pt,partopsep=0pt,parsep=\parskip,topsep=5pt}

\begin{document}

\title{\textit{PVMark}: Enabling Public Verifiability for LLM Watermarking Schemes
	}

\author{Haohua~Duan,
	Liyao~Xiang*,~\IEEEmembership{Member,~IEEE},
	and~Xin~Zhang
	\IEEEcompsocitemizethanks{\IEEEcompsocthanksitem H. Duan, L. Xiang (xiangliyao08@sjtu.edu.cn, the corresponding author) are with Shanghai Jiao Tong University, Shanghai 200240, China. E-mail: \{duanhaohua, xiangliyao08\}@sjtu.edu.cn. X. Zhang is with Ant Group. E-mail: evan.zx@antgroup.com.

}}

\maketitle

\begin{abstract}
Watermarking schemes for large language models (LLMs) have been proposed to identify the source of the generated text, mitigating the potential threats emerged from model theft. However, current watermarking solutions hardly resolve the trust issue: the non-public watermark detection cannot prove itself faithfully conducting the detection. We observe that it is attributed to the secret key mostly used in the watermark detection --- it cannot be public, or the adversary may launch removal attacks provided the key; nor can it be private, or the watermarking detection is opaque to the public. To resolve the dilemma, we propose PVMark, a plugin based on zero-knowledge proof (ZKP), enabling the watermark detection process to be publicly verifiable by third parties without disclosing any secret key. PVMark hinges upon the proof of `correct execution' of watermark detection on which a set of ZKP constraints are built, including mapping, random number generation, comparison, and summation. We implement multiple variants of PVMark in {\tt Python}, {\tt Rust} and {\tt Circom}, covering combinations of three watermarking schemes, three hash functions, and four ZKP protocols, to show our approach effectively works under a variety of circumstances. By experimental results, PVMark efficiently enables public verifiability on the state-of-the-art LLM watermarking schemes yet without compromising the watermarking performance, promising to be deployed in practice.
\end{abstract}


\section{Introduction}
\IEEEPARstart{I}n recent years, large language models (LLMs) have made significant impact to the community by generating high-quality texts akin to human-written ones. However, this advancement also brings many real-world threats such as the spread of generated fake news \cite{pan2023riskmisinformationpollutionlarge}, misuse of LLMs in academic misconduct \cite{vasilatos2023howkgptinvestigatingdetectionchatgptgenerated}, etc. To achieve responsible AI, it is important to trace the provenance of the generated texts by LLMs.

\begin{figure}[tb]
	\centering
	\includegraphics[width=0.98\linewidth]{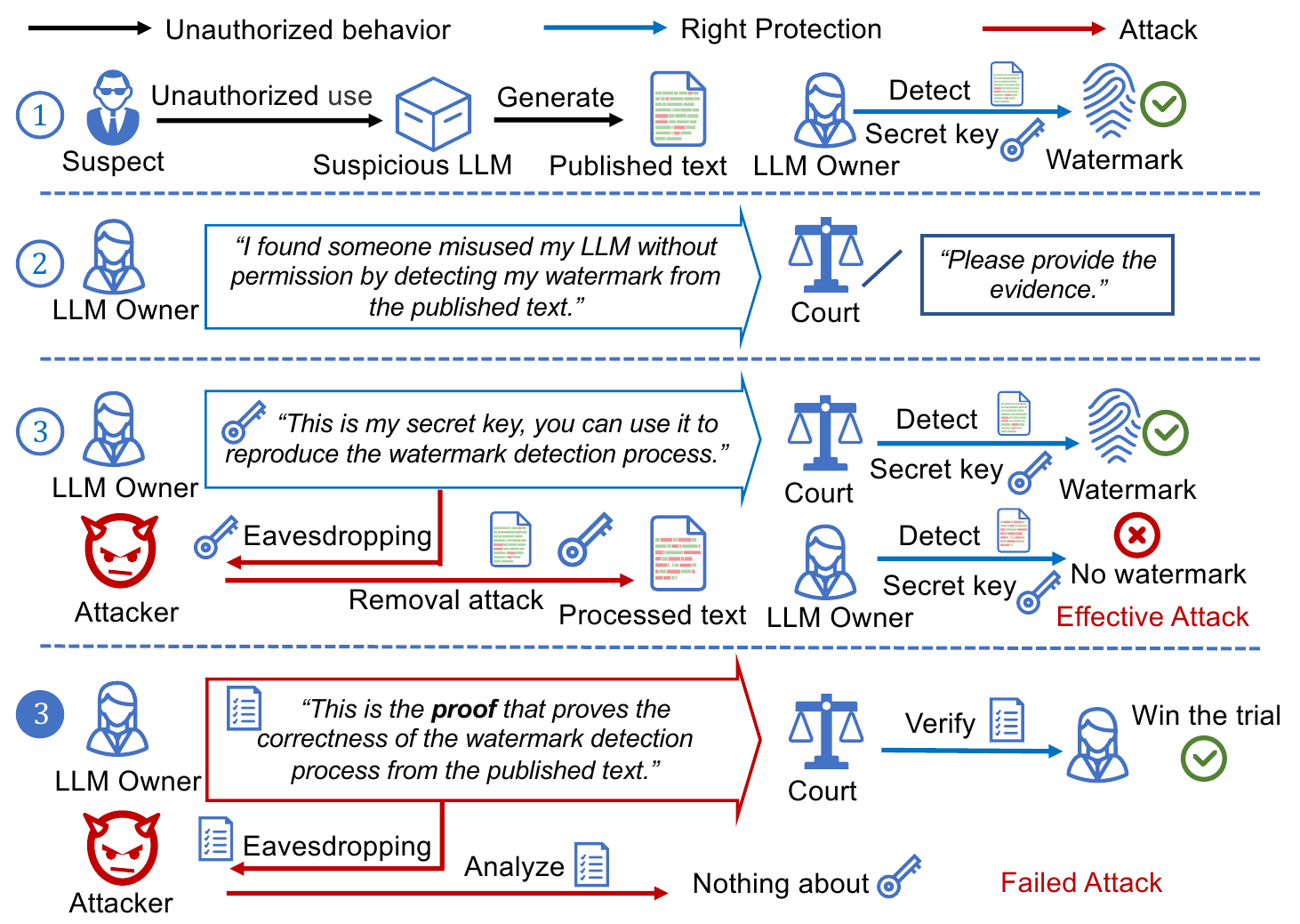}
	\caption{From top to bottom: \ding{192} The LLM owner detects unauthorized use of the LLM by watermark detection. \ding{193} The LLM owner sues the suspected model and the the court requests evidence. \ding{194} The LLM owner's secret key gets exposed in answering the court's request. \ding{204} The LLM owner provides the proof as evidence under \textit{PVMark} thus preventing the attack.}
	\label{fig::dilemma}
	\vspace{-0.5cm}
\end{figure}

Towards this end, various watermarking schemes \cite{kirchenbauer2023watermark,dathathri2024scalable,lee2024wrotecodewatermarkingcode,wang2024codablewatermarkinginjectingmultibits,yoo2024advancingidentificationmultibitwatermark,kirchenbauer2024reliabilitywatermarkslargelanguage,zhao2023provablerobustwatermarkingaigenerated,hu2023unbiasedwatermarklargelanguage,wu2024resilientaccessibledistributionpreservingwatermark,qu2025provablyrobustmultibitwatermarking} are proposed for LLMs. The mainstream of the watermarking schemes embeds the watermark by altering the probability distribution of the generated tokens depending on the context and owner's secret key. The watermark is later extracted by detecting the statistical significance in the word distribution of the suspected text. However, such watermarking schemes face a trust crisis in practical applications. As shown in Figure~\ref{fig::dilemma}, when the LLM owner proclaims its right against unauthorized use, it may face questions from a third party, e.g., the court, about the legitimacy of the watermark, due to the opaque watermark detection involving a secret key. In such a case, the owner has to risk its secret key to third-party detection, while the third party may be non-confidential allowing an eavesdropper to acquire the secret key, threatening the validity of the watermark. The root cause of this problem is the use of a secret key which cannot be public --- otherwise putting the watermarked model under the threat of removal or ambiguity attacks; nor private --- for hardly convincing others that the detection is legitimate. Overall, LLM watermarking schemes lack \textit{public verifiability}.

Previous efforts have been devoted to resolve the issue. Fairoze \emph{et al.} attempt to use a private key for watermark embedding and a public key for detection \cite{fairoze2024publiclydetectablewatermarkinglanguagemodels}. However, the detection relies much on extracted features of the generated texts, which can be exploited by adversaries. Liu \emph{et al.} address the problem by training two separate neural networks as watermark embedding and detection modules respectively \cite{liu2023unforgeable}, but the exposure of the detection network still enables the adversary to train a reverse network to remove the watermark. Kirchenbauer \emph{et al.} propose a private detection scheme, allowing only authorized parties to access the detection APIs \cite{kirchenbauer2023watermark}, yet without convincing others of the faithful execution of APIs.

To resolve the dilemma, we design a plugin named \textit{PVMark}, which augments the state-of-the-art LLM watermarking schemes with \textit{public verifiability} while preserving the original schemes' properties of effectiveness, fidelity, and robustness. Specifically, without revealing the secret key, the owner or an authorized party can perform a publicly verifiable detection to convince others that the detection is faithfully done, thus preventing any attack raised by theft of secret key.

\textit{PVMark} achieves public verifiability by integrating zero knowledge proof (ZKP) into watermark detection. ZKP is a protocol in which the prover can convince the verifier that a given statement is true, without conveying to the verifier any information beyond the fact that the statement is true. It is aligned with our goal of verifying the watermark detection is correctly done yet without revealing the secret key. 

To instantiate the idea, we select three representative LLM watermarking schemes, i.e., KGW~\cite{kirchenbauer2023watermark}, SynthID-Text~\cite{dathathri2024scalable}, and Segment-Watermark~\cite{qu2025provablyrobustmultibitwatermarking}, to implement \textit{PVMark}. We found these schemes share a common token-sampling-alteration framework: at each sampling round, all candidate tokens are randomly divided into different groups based on the secret key and the watermark. Then tokens in the `green' group are given a slightly higher probability to be sampled as the output. Hence the statistical attributes of the output can be used to extract the watermark.

Our core idea is to provide evidence that the statistical results of green tokens in the produced text are faithfully collected without exposing the secret key, thereby demonstrating that the text indeed contains watermarks. Hence two criteria need to satisfy: the watermark is extracted from the text, and the proof of the detection passes verification. In applying ZKP to watermarking, we mainly resovle two technical challenges: 1) how to model the watermark detection by circuits on which ZKP protocols run, and 2) how to minimize the overhead introduced by ZKP thereby improving detection efficiency.

To tackle challenge 1, we decompose the watermark detection into four components for reconstruction — mapping table check, hashing, comparison, and summation, replacing the original components by ZKP-friendly ones. For example, the pseudorandom function (PRF) instantiated by a modular operation-based linear congruential generator is inefficient for ZKP, which is substituted by cryptographic hash functions. The identification of `green' tokens is reconfigured as comparing the generated random numbers against a fixed threshold. We show that the reconstruction of the watermarking schemes not only preserves the watermark properties, but also tailor them to ZKP protocols.

Apart from reconstruction, we propose multiple approaches to improve the efficiency of \textit{PVMark}~(addressing challenge 2).
Since hash functions are heavily used in detection, we combine them into three-to-one hash functions for more succinct constraints. Observing the watermark detection is a repeated procedure on a sequence of tokens, we treat the detection on a subsequence of tokens as an instance and reduce checking of multiple instances into checking one by folding~\cite{kothapalli2022nova}. Hence we reduce the ZKP overhead for proving multiple instances into proving one final instance and the folding steps, significantly mitigating the cost of \textit{PVMark}.

Highlights of our contributions are as follows. We propose \textit{PVMark} that enables \textit{public verifiability} on zero- and multi-bit LLM watermarking schemes, achieving a compromise between public detection and authorized detection, resolving the trust issue in watermarking detection. \textit{PVMark} is built upon ZKP techniques and incorporates application-specific optimizations. Implementation-wise, we realize multiple variants of \textit{PVMark} on {\tt Python}, {\tt Rust} and {\tt Circom}, including different combinations of three LLM watermarking schemes, three hash functions, and four ZKP protocols. The security analysis and experimental results have demonstrated that \textit{PVMark} is applicable to be deployed in LLM watermarking service with low costs for public verification.

\section{Related Work}

Currently, multiple schemes are proposed for LLM watermarking. One is retrieval-based schemes~\cite{krishna2024paraphrasing} that require accessing and storing all the texts generated during the interaction process of LLM. This largely destroys the privacy of users using LLM. The second is post hoc detection schemes~\cite{mitchell2023detectgpt,verma-etal-2024-ghostbuster,hans2024spottingllmsbinocularszeroshot} that usually train a machine learning model to distinguish human-written text from LLM-generated text. These schemes require a lot of computing resources and the performance is not good enough~\cite{elkhatat2023evaluating}. The third is represented by KGW scheme~\cite{kirchenbauer2023watermark}. The high-level idea of KGW is to divide the vocabulary into green and red token lists, and modify logits to shift the distribution of candidate tokens toward the green lists. This approach allows watermark embedding without requiring any LLM fine-tuning, making it general and less costly to deploy.

Many watermarking schemes based on probability shifting have been proposed subsequently. Lee \emph{et al.}~\cite{lee2024wrotecodewatermarkingcode}  and Wang \emph{et al.}~\cite{wang2024codablewatermarkinginjectingmultibits} reduce the impact of logits modification on text quality by only watermarking high-entropy tokens. Yoo \emph{et al.} propose to assign different message bits to distinct positions within the text to increase the payload of the watermark in~\cite{yoo2024advancingidentificationmultibitwatermark}. Qu \emph{et al.} propose using dynamic programming algorithms to allocate tokens to each position of the message to be embedded in a balanced manner to improve the accuracy of watermark detection. For robustness, Kirchenbauer \emph{et al.} study the effects of different hash strategies on watermarking robustness in~\cite{kirchenbauer2024reliabilitywatermarkslargelanguage}. Zhao \emph{et al.} demonstrate that using a globally fixed vocabulary partitioning scheme provides better resistance against removal attacks in~\cite{zhao2023provablerobustwatermarkingaigenerated}. Hu \emph{et al.}~\cite{hu2023unbiasedwatermarklargelanguage} and Wu \emph{et al.} ~\cite{wu2024resilientaccessibledistributionpreservingwatermark} propose an unbiased weighting method to improve the imperceptibility of KGW. Dathathri \emph{et al.} propose Tournament sampling algorithm to improve watermark detection rate while preserving text quality in~\cite{dathathri2024scalable}.

 However, the aforementioned schemes are only suitable for private watermark detection scenarios and cannot be applied to public detection, as exposing the detector would face various threats. 

To address the public verifiability of watermark, Fairoze \emph{et al.} utilize the asymmetric encryption to embed the watermark using a private key and detect it by a public key~\cite{fairoze2024publiclydetectablewatermarkinglanguagemodels}. However, this approach relies on text features for watermark detection, which can also be exploited by adversaries for spoofing attacks. Liu \emph{et al.} propose training two additional neural network models in~\cite{liu2023unforgeable}, one for watermark embedding and another for watermark detection. But, when the window size, which decides the partition of vocabulary for the current token, is relatively small, the method would suffer from reverse engineering attack, i.e., reverse training of the embedding network from the detection network, threatening the robustness of the system.

\section{Preliminary}

\subsection{Zero Knowledge Proof}
\label{sec::pre_zkp}
Zero knowledge proofs (ZKPs) is a tool whereby a computationally unbounded party (the prover) can prove to another party (the verifier) that a given statement is true, while avoiding conveying any additional information apart from the fact that the statement is indeed true. ZKP ensures that if the prover has the correct private input $w$ and public input $u$, it can convince the verifier of the statement's validity with a very high probability (completeness property), without revealing anything about $w$ (zero-knowledge property). Conversely, if the prover does not have $w$, the verifier will reject the proof with a very high probability (soundness property). The formal definitions of zero knowledge arguments can be found in supplementary file.

In the context of ZKPs, an arithmetic circuit is a system of equations that models a problem and on which ZKP protocols run to verify. Common circuits include GKR-based ones, rank-1 constraint system (R1CS), PLONKish arithmetization, etc. For simplicity, we use PLONKish as an example type of circuit to introduce our system. From a high-level perspective, PLONKish arithmetization encompasses the numerical values of both public and private inputs, along with equations that describe the relationships among these values, denoted as constraints. The form of constraints used by the standard PLONKish is as follows:
\begin{equation}\label{eq:standard_plonkish}
	s_L \cdot x_L + s_R \cdot x_R + s_O \cdot x_O + s_M \cdot x_L \cdot x_R + s_C \cdot c = 0,
\end{equation}
where $s_L, s_R, s_O, s_M, s_C \in \mathbb{F}$ are selectors, and $x_L, x_R, x_O, c \in \mathbb{F}$ denotes the left input, the right input, the output and the constant value, respectively. %

\textbf{Recursive ZKP.} Nova \cite{kothapalli2022nova} is an effective recursive zero-knowledge protocol from folding schemes to realize incrementally verifiable computation (IVC). IVC refers to recursively verifiable computations of the form
\begin{equation}
	\label{eq:ivc_form}
	u_{\eta}=F(w_{\eta - 1}, F(w_{\eta - 2}, F(\ldots F(w_1, F(w_0, u_0)))))
\end{equation}
where $F, w, u$ and $\eta$ denote the computation function, private input, public input and the number of computations, respectively. Nova can fold two relaxed R1CS instances into one and the relaxed R1CS can be expressed as follows:
\begin{equation}
	\label{eq:relaxed_r1cs}
	(\mathsf{A} \cdot \mathsf{Z}) \odot (\mathsf{B} \cdot \mathsf{Z}) = \mu (\mathsf{C} \cdot \mathsf{Z}) + \nu
\end{equation}
where $\mathsf{A}, \mathsf{B}$, $\mathsf{C}$ are matrices containing ZKP circuits information, $\mu$ is a scalar and $\nu$ is used to store redundant items. The $\odot$ denotes the hadamard product. $\mathsf{Z}=[u,w,1]$ is the instance vector including the witness $w$, public input $u$ and the constant $1$. Two relaxed R1CS instance ($\mathsf{Z}_1, \mu_1, \nu_1$) and ($\mathsf{Z}_2, \mu_2, \nu_2$) can be folded into one instance ($\mathsf{Z}, \mu, \nu$) as follows:
	\begin{align}
			\nu =& \nu_{1} + r \cdot (\mathsf{A} \mathsf{Z}_{1} \odot \mathsf{B} \mathsf{Z}_{2} + \mathsf{A} \mathsf{Z}_{2} \odot \mathsf{B} \mathsf{Z}_{1} \notag \\
			&- \mu_{1} \cdot \mathsf{C} \mathsf{Z}_{2} - \mu_{2} \cdot \mathsf{C} \mathsf{Z}_{1}) + r^{2} \cdot \nu_{2} , 	\label{eq:fold1} \\
			\mu =& \mu_{1} + r \cdot \mu_{2}, 	\label{eq:fold2} \\
			\mathsf{Z} =& \mathsf{Z}_1 + r \cdot \mathsf{Z}_2 ,	\label{eq:fold3}
	\end{align}
where $r$ denotes randomness. It suffices to provide proofs for the correctness of Eq.~\eqref{eq:fold1}, ~\eqref{eq:fold2}, ~\eqref{eq:fold3}, and the folded instance, eliminating the need for proofs of the original two instances. With careful design, savings are achieved if the constraints for a single fold are fewer than those for an individual instance. %

\subsection{Language Model Watermarking}
\label{sec.llm_watermark}
A language model (LM) is a type of machine learning model trained to predict the next most appropriate word to fill in a blank space in a sentence or phrase, based on the given text. Given an input prompt $\boldsymbol{x} = [x_0, \ldots, x_{m - 1}]$ of length $m$, a language model $\mathcal{M}$ is a function that iteratively computes the logit scores $l_j^{(i)}$ ($0 \leq i \leq n - 1, j \in |V|$) of the next $n$ tokens over the vocabulary $V$. These logits are transformed into a probability distribution $\boldsymbol{p}^{(i)}$ via the softmax and the LM $\mathcal{M}$ samples the next $n$ tokens $\boldsymbol{y} = [y_0, \ldots, y_{n - 1}]$ from $\boldsymbol{p}^{(i)}$. We overload the syntax $y_i$ to denote the index of the corresponding token in the vocabulary without causing confusion.

Language model watermarking is a technique that allows the model owner to embed identifiable messages known as watermarks within the sequence $\boldsymbol{y}$ generated by a watermarked LM $\hat{\mathcal{M}}$. We give the formal definition of language model watermarking in the supplementary file due to space limitation.

\definecolor{darkyellow}{RGB}{220, 120, 0}

\begin{table*}[h]
	\centering %
	\renewcommand{\arraystretch}{1.25}
	\caption{The watermark embedding process of zero-bit and multi-bit LLM watermarking schemes. Steps requiring adjustment for PVMark are marked in brackets. $\text{PRF}(\cdot)$ denotes a pseudo-random function.}
	\label{tab:watermark_embed}
	\resizebox{\linewidth}{!}{
		\begin{tabular}{m{0.07\linewidth}|m{0.25\linewidth}|m{0.568\linewidth} m{0.15\linewidth}} %
			\noalign{\hrule height 1.0pt}
			& \textbf{Step 1}: Random seed generation.
			& \multicolumn{1}{m{0.25\linewidth}|}{\textbf{Step 2}:  Vocabulary partitioning.}
			& \textbf{Step 3}: Probability shifting and sampling. \\ 
			\hline  
			\textbf{KGW}~\cite{kirchenbauer2023watermark}            & Compute random seed $sd_i \gets \text{PRF}(sk, \boldsymbol{y}_{\psi})$. $\Rightarrow$ [Adjustment for PVMark: Compute random seed $sd_i \gets \text{Hash}(sk, \boldsymbol{y}_{\psi})$.]
			& \multicolumn{1}{m{0.668\linewidth}|}{ a) Assign random number $g^{(y_j)} = \text{PRF}(sd_i, y_j)$ to each token $y_j \in V$; $\Rightarrow$ [Adjustment for PVMark: Assign random number $g^{(y_j)} = \text{Hash}(sd_i, y_j)$ to each token $y_j \in V$;] \newline
				b) Sort $V$ by all the random numbers $g^{(y_1)}, \ldots, g^{(y_{|V|})}$ corresponding to token $y_1, \dots, y_{|V|} \in V$ and designate the top $\gamma|V|, \gamma \in (0,1)$ tokens in the sorted $V$ as green list $G$. $\Rightarrow$ [Adjustment for PVMark: Partition the all candidate tokens as follows:
				\[ \text{Candidate token} \ y_j \in 
				\begin{cases}
					G, & g^{(y_j)} < \gamma \cdot |\mathbb{F}| \\
					R, & g^{(y_j)} \geq \gamma \cdot |\mathbb{F}| 
				\end{cases}
				\]
				for $j \in |V|$, where $R$ and $\mathbb{F}$ denotes the red list and finite field, respectively.
				]
			}

			& Add a bias $\delta$ to logits of all tokens in $G$ and compute the new probability distribution $\hat{{\boldsymbol{p}}}^{(i)}$ using new logits over $V$. Sample $y_i$ from $\hat{{\boldsymbol{p}}}^{(i)}$ and append $y_i$ to $\boldsymbol{y}$.  \\ \cline{1-1} \cline{3-4}
			\textbf{SynthID-Text}~\cite{dathathri2024scalable}      
			&  
			& \multicolumn{2}{m{0.718\linewidth}}{
				\begin{multicols}{2}
					\begin{algorithmic}
						\STATE Generate top K candidate set $V$ for this round and assign $\xi$ random numbers $g_k^{(y_j)} = \text{PRF}(sd_i, y_j, k)$ to each token $y_j \in V$ for $k \in \{1, \ldots, \xi \}$; $\Rightarrow$ [Adjustment for PVMark: Generate top $K = 2^\xi$ candidate set $V$ for this round and assign $\xi$ random numbers to each token $y_j \in V$ by
						$$g_k^{(y_j)} = \begin{cases}
							1, & \text{Hash}(sd_i, y_j, k) < |\mathbb{F}| / 2, \\
							0, & \text{Hash}(sd_i, y_j, k) \geq |\mathbb{F}| / 2, 
						\end{cases}$$ 						for $k \in \{1, ..., \xi \}$;
						]
						\STATE For each $y_j \in V$, set $y_j^{(0)} \gets y_j$.
						\FOR{$k = 1, \ldots, \xi$}
						\FOR{$j = 0, \ldots, 2^{\xi - k} - 1$}
						\STATE $J = [y_{2j}^{(k-1)}, y_{2j+1}^{(k-1)}]$ (may contain repeats).
						\STATE $J^{\ast} = [y \in J ~ s.t. ~ g_k^{(y)}= \max_{y^{\prime} \in J} g_k^{(y^{\prime})}]$ (may contain repeats).
						\STATE Sample $y_j^{(k)}$ from the uniform distribution of $J^{\ast}$.
						\ENDFOR
						\ENDFOR
						\STATE Append $y_0^{\xi}$ to $\boldsymbol{y}$.
					\end{algorithmic}
				\end{multicols}
			} \\ \cline{1-4}
			\textbf{Segment-Watermark} \cite{qu2025provablyrobustmultibitwatermarking} 
			& a) Find position index $p_i = M[\boldsymbol{y}_{\psi}]$;
			b) Compute random seed $sd_i \gets \text{PRF}(sk, \boldsymbol{y}_{\psi}, \boldsymbol{msg}[p_i])$. $\Rightarrow$ [Adjustment for PVMark: Compute random seed $sd_i \gets \text{Hash}(sk, \boldsymbol{y}_{\psi}, \boldsymbol{msg}[p_i])$.]
			& \multicolumn{1}{m{0.15\linewidth}|}{The same as \textbf{KGW}. }
			& The same as \textbf{KGW}.  \\ 
			\noalign{\hrule height 1.0pt}  
		\end{tabular}
	}
\end{table*}

\begin{table*}[h]
	\renewcommand{\arraystretch}{1.25}
	\centering
	\caption{The watermark detection process of zero-bit and multi-bit LLM watermarking schemes. Steps requiring adjustment for PVMark are marked in brackets.}
	\label{tab:watermark_detection}
	\resizebox{\linewidth}{!}{
			\begin{tabular}{m{0.07\linewidth}|m{0.20\linewidth}|m{0.46\linewidth}|m{0.26\linewidth}}
				\noalign{\hrule height 1.0pt}
				& \textbf{Step 1}: Compute random seed 
				& \textbf{Step 2}: Check and count green tokens
				& \textbf{Step 3}: Compute score / message \\ 
				\hline  %
				\textbf{KGW}~\cite{kirchenbauer2023watermark}            
				& Compute random seed $sd_i \gets \text{PRF}(sk, \boldsymbol{y}_{\psi})$. $\Rightarrow$ [Adjustment for PVMark: Compute random seed $sd_i \gets \text{Hash}(sk, \boldsymbol{y}_{\psi})$.]
				& a) Similar to Step 2 of the watermark embedding process, assign a random number to each token in the vocabulary using $sd_i$ and reorder the vocabulary to find $G_i$; $\Rightarrow$ [Adjustment for PVMark: Compute the random number $g^{y_i} = \text{Hash}(sd_i, y_i)$ corresponding to the current token $y_i$;]
				
				b) Count the number of green tokens $|\boldsymbol{y}|_{G} \mathrel{+}= \mathds{1}(y_i \in G_i)$ where $\mathds{1}(\cdot)$ is the indicator function. $\Rightarrow$ [Adjustment for PVMark: Count the number of green tokens $|\boldsymbol{y}|_{G} \mathrel{+}= \mathds{1}(g^{y_i} < \gamma \cdot |\mathbb{F}|)$.]
				& $\text{Score}(\boldsymbol{y}) = (|\boldsymbol{y}|_{G} - \gamma n) / \sqrt{n \gamma(1-\gamma)}$ 
				\\ \cline{1-1} \cline{3-4}
				\textbf{SynthID-Text}~\cite{dathathri2024scalable}      
				&  
				& a) Compute $\xi$ random number $g_k^{y_i}$ corresponding to the current token $y_i$ where $k \in \{1, \ldots, \xi \}$; $\Rightarrow$ [Adjustment for PVMark: Compute $\xi$ random number corresponding to the current token $y_i$ by
				$$g_k^{(y_i)} = \begin{cases}
					1, & \text{Hash}(sd_i, y_i, k) \textless |\mathbb{F}| / 2, \\
					0, & \text{Hash}(sd_i, y_i, k) \geq |\mathbb{F}| / 2, 
				\end{cases}$$ 					for $k \in \{1, ..., \xi \}$;
				]
				
				b) Update $S_g$: $S_g \mathrel{+}= \sum_{k=1}^{\xi} g_k$.
				& $\text{Score}(\boldsymbol{y}) = \frac{S_g}{(n- \psi ) \cdot \xi}$ 
				\\ \cline{1-4}
				\textbf{Segment-Watermark}\cite{qu2025provablyrobustmultibitwatermarking} 
				& a) Find position index $p_i = M[\boldsymbol{y}_{\psi}]$;
				
				b) Compute all possible random seeds $sd_i^j \gets \text{PRF}(sk, \boldsymbol{y}_{\psi}, j)$ for $j = 0, 1, \ldots, 2^{\hat{m}}-1$. $\Rightarrow$ [Adjustment for PVMark: Compute all possible random seeds $sd_i^j \gets \text{Hash}(sk, \boldsymbol{y}_{\psi}, j)$ for $j = 0, 1, \ldots, 2^{\hat{m}}-1$.]
				& \begin{algorithmic}
					\FOR{$j = 0, 1, \ldots, 2^{\hat{m}}-1$}
					\STATE Assign a random number to each token in the vocabulary using $sd_i^j$ and reorder the vocabulary to find $G_i^j$; $\Rightarrow$ [Adjustment for PVMark: Compute the random number $g^{y_i} = \text{Hash}(sd_i^j, y_i)$ corresponding to the current token $y_i$;]
					\STATE Update COUNT: $\text{COUNT}[p_i][j] \mathrel{+}= \mathds{1}(y_i \in G_i)$.  $\Rightarrow$ [Adjustment for PVMark: Update COUNT: $\text{COUNT}[p_i][j] \mathrel{+}= \mathds{1}(g^{y_i} < \gamma \cdot |\mathbb{F}|)$.]
					\ENDFOR
				\end{algorithmic}
				&
				$$\boldsymbol{msg}[p] \gets \arg\max_{j \in [0, 2^{\hat{m}-1} ]} \text{COUNT}[p][j]$$ for $p = 0, 1, \ldots, \hat{n} - 1$.
				\\
				\noalign{\hrule height 1.0pt}  %
			\end{tabular}
		}
\vspace{-0.4cm}	
	\end{table*}

\textbf{Common LLM watermarking frameworks.}
Existing LLM watermarking schemes based on probability shifting can be categorized into zero-bit watermarking schemes~\cite{kirchenbauer2023watermark,lee2024wrotecodewatermarkingcode,wang2024codablewatermarkinginjectingmultibits,yoo2024advancingidentificationmultibitwatermark,kirchenbauer2024reliabilitywatermarkslargelanguage,zhao2023provablerobustwatermarkingaigenerated,hu2023unbiasedwatermarklargelanguage,wu2024resilientaccessibledistributionpreservingwatermark,dathathri2024scalable,zhao2024provable} and multi-bit watermarking schemes~\cite{qu2025provablyrobustmultibitwatermarking} based on their capability to embed messages into LLM-generated text. The former type only injects features to differentiate AI-generated text from human-produced text without embedding any explicit information, whereas the latter can embed multi-bit metadata — such as a model owner’s identity — into text for more precise identification and traceability.

Despite differences in technical details, these two categories of probability-shifting-based watermarking schemes share a common high-level framework. We select representative works in zero-bit watermarking schemes, namely KGW~\cite{kirchenbauer2023watermark} and SynthID-Text~\cite{dathathri2024scalable}, as well as Segment-Watermark~\cite{qu2025provablyrobustmultibitwatermarking} in multi-bit watermarking schemes to showcase these watermarking schemes.

The watermark embedding process is summarized in Table~\ref{tab:watermark_embed} and can be divided into three steps: random seed generation, vocabulary partitioning, and probability shifting and sampling. Particularly for SynthID-Text, the latter two steps are indistinguishable. 

In each round $i$, the first step is to generate a random seed $sd_i$ based on the secret key $sk$ and contextual information $\boldsymbol{y}_{\psi}$ where $\psi$ denotes the recent context window width. This step in multi-bit watermarking additionally requires the mapping of tokens into positions ($p$) and the position index is used to pick the corresponding message bit ($\boldsymbol{msg}[p]$) to generate the random seed. The seed $sd_i$ is then employed to partition candidate tokens into groups for the second step. In Step 3, the logits of tokens in the selected groups are adjusted accordingly by means of adding a bias to increase the probability of being sampled. In essence, this probability-shifting mechanism inclines the LLM to preferentially output tokens from the selected group, thereby embedding watermark information into the LLM-generated text. In particular for SynthID-Text, the probability shifting is applied implicitly through Tournament sampling where every two candidates compare their $g$ values to enter the next-tier competition until the final winner is selected. It has been proved that the Tournament sampling in fact raises the probability of the winner and lowers the probability of the other through each round of competition.

The watermark detection processes for zero-bit and multi-bit schemes exhibit subtle distinctions, shown in Table~\ref{tab:watermark_detection}. For zero-bit schemes, the random seed derived from the secret key $sk$ and context $\boldsymbol{y}_{\psi}$ is used to determine the token grouping for each round. The number of tokens in the given text $\boldsymbol{y}$ that belong to the selected group is counted as $|\boldsymbol{y}|_{G}$. A watermark score $\text{Score}(\boldsymbol{y})$ is computed based on $|\boldsymbol{y}|_{G}$, and this score is compared against a threshold $\tau$ to identify the being of the watermark. In contrast, multi-bit schemes involve iterating over all possible watermark message values in calculating random seeds, alongside the secret key and context. By analyzing token group statistics associated with different watermark message values, the detection algorithm selects the highest one as the decoded message.

\section{Problem Statement and PVMark Overview}
\label{sec:problem}
In this section, we point out the threats that the common LLM watermarking frameworks face in ownership verification or generated content provenance tracing. We formally define the problem and then give a solution overview.

\textbf{Threat model.}
In ownership verification and provenance tracing, the watermarking frameworks involve a watermark embedding party, usually the model owner, and a watermark detection party. The detection party extracts the watermark from the watermarked text as evidence of the text origins. In previous works \cite{kirchenbauer2023watermark,lee2024wrotecodewatermarkingcode,wang2024codablewatermarkinginjectingmultibits,yoo2024advancingidentificationmultibitwatermark,kirchenbauer2024reliabilitywatermarkslargelanguage,zhao2023provablerobustwatermarkingaigenerated,hu2023unbiasedwatermarklargelanguage,wu2024resilientaccessibledistributionpreservingwatermark,dathathri2024scalable,zhao2024provable,qu2025provablyrobustmultibitwatermarking}, the detection parties are mostly the embedding parties, as they own the detection key for watermark detection. However, the confidential detection process would \textit{reduce its credibility} to the public. If the detection party is acted by a third party, the party has to keep the detection key confidential. The leakage of the detection key to a malicious party would tempt an adaptive attack to remove the watermark from the watermarked content, known as \textit{removal attacks}. This implies that the aforementioned LLM watermarking frameworks lack \textit{public verifiability}.

\textbf{Motivation.} 
We observe that the lack of public verifiability is due to the key $sk$ used in the embedding and the detection phases. The $sk$ is crucial to be kept secret as its exposure would leak the vocabulary partition for watermark embedding, allowing adversaries to identify `green tokens,' and thus intentionally replacing them to remove the watermark.

Our intuition is that, $sk$ cannot be made public while the watermark detection should be transparent. To resolve the dilemma, we propose a plugin \textit{PVMark} that enables public verification of LLM watermarking without leaking $sk$: it provides a correctness proof of the watermark detection to any verifier, which circumvents the direct exposure of the detection process. Specifically, by integrating ZKP into watermarking detection, \textit{PVMark} can hide $sk$ (zero-knowledge property) while providing evidence that the detection is correctly performed by the relevant party (completeness property). Meanwhile, the probability of cheating to pass verification is negligible (soundness property). Even though the proof only serves the detection, the embedding procedure requires adaptation to the proof.

\textbf{Problem statement.} 
Formally, four parties are involved in \textit{PVMark}: (1) \textit{watermark embedding party} $\mathcal{E}$ who embeds the watermark into $\mathcal{M}$ to obtain $\hat{\mathcal{M}}$; (2) \textit{watermark detection party} $\mathcal{D}$ who has access to the watermark detector, $sk$ corresponding to $\hat{\mathcal{M}}$, and a text sequence $\boldsymbol{y}$. $\mathcal{D}$ extracts watermark from $\boldsymbol{y}$ while acting as the \textit{prover} to prove the claim that `$\boldsymbol{y}$ indeed contains watermark, and the watermark information is $\boldsymbol{msg}$ (for multi-bit watermarking);' (3) \textit{verifier} $\mathcal{V}$ who could be any party responsible to verify $\mathcal{D}$'s claim. $\mathcal{V}$ is honest in executing the proof verification while being curious about $sk$ and other credentials that might be contained in the proof provided by $\mathcal{D}$; (4) \textit{adversary} $\mathcal{A}$ who maliciously exploits $\hat{\mathcal{M}}$ to generate unauthorized text with watermark removed.

\textbf{Use case.} 
\textit{Ownership verification.} The model owner $\mathcal{E}$ trains a language model $\mathcal{M}$ and embeds its watermark to get $\hat{\mathcal{M}}$. If any unauthorized party is suspected to illegally possess $\hat{\mathcal{M}}$ and generate text $\boldsymbol{y}$ by it, the watermark detection party $\mathcal{D}$, typically the owner or an authorized party, could provide the evidence by detecting the watermark from $\boldsymbol{y}$, generating the corresponding correctness proof and submitting it to the verifier $\mathcal{V}$. It is confirmed that the unauthorized party illegally possesses $\hat{\mathcal{M}}$, should the watermark be detected and the proof be verified.

\textit{Provenance tracing.} The owner $\mathcal{E}$ generates content $\boldsymbol{y}$ with watermark embedded. Someone raises concern on $\boldsymbol{y}$ and tracks the origin of the text. The authorized party $\mathcal{D}$ detects watermark belonging to $\mathcal{E}$ and provides evidence to $\mathcal{V}$ that the detection is faithfully done. $\mathcal{E}$ cannot deny the results supported by evidence.

 In both use cases, the embedding and detection algorithms are publicly known, except for the secret key $sk$. $\mathcal{D}$ uses $sk$ in detection and proof generation without revealing it.
 
\textbf{Overview.} 
Formally, we define \textit{PVMark} as follow:
\begin{definition}
	PVMark is a plugin that enables public verifiability on a language model watermarking scheme. It consists of a triple of setup, embedding and detection algorithms: 
	\begin{itemize}[leftmargin=*]
		\item Setup($sk$) $\rightarrow \Upsilon$: $\mathcal{E}$ makes a commitment $\Upsilon$ to a secret key $sk$ and publishes $\Upsilon$.
		\item Watermark embedding WE($sk$, $\mathcal{M}$) $\rightarrow \hat{\mathcal{M}}$: $\mathcal{E}$ uses the secret key $sk$ to produce a watermarked model $\hat{\mathcal{M}}$.
		\item Watermark detection WD($sk$, $\boldsymbol{y}$) $\rightarrow \{0, 1\}$ or $\boldsymbol{msg}, \Pi$: $\mathcal{D}$ takes the secret key $sk$ associated with $\hat{\mathcal{M}}$ and a text sequence $\boldsymbol{y}$ as inputs. For zero-bit watermarking schemes, it detects whether $\boldsymbol{y}$ contains watermark and returns 1 if present, otherwise 0; for multi-bit watermarking schemes, it detects and returns the watermark information $\boldsymbol{msg}$ embedded in $\boldsymbol{y}$. Then $\mathcal{D}$ generates a correctness proof $\Pi$ for the detection.
	\end{itemize}
\end{definition}

In previous work~\cite{kirchenbauer2023watermark,lee2024wrotecodewatermarkingcode,wang2024codablewatermarkinginjectingmultibits,yoo2024advancingidentificationmultibitwatermark,kirchenbauer2024reliabilitywatermarkslargelanguage,zhao2023provablerobustwatermarkingaigenerated,hu2023unbiasedwatermarklargelanguage,wu2024resilientaccessibledistributionpreservingwatermark,dathathri2024scalable,zhao2024provable,qu2025provablyrobustmultibitwatermarking}, the properties of watermarking schemes have predominantly focused on \textit{effectiveness and fidelity}, meaning the watermark should be embedded and detected with high accuracy while maintaining decent text quality, and \textit{robustness}, which hinders adversaries from removing the watermark from watermarked text or forging the watermark into unwatermarked text. We claim that PVMark can augment watermarking schemes with an additional property of \textit{public verifiability} without undermining the aforementioned attributes --- specifically, the property allows any third party to check the integrity of the watermark detection result.

It is noteworthy that Pang \emph{et al.} point out an inherent conflict between robustness and public detectability for watermarking systems in \cite{pang2024freelunchllmwatermarking}. In contrast, \textit{PVMark} effectively achieves robustness and public verifiability at the same time. \textit{PVMark} is not a publicly detectable scheme, but its watermarking detection process can be verified publicly, adding credibility to the detection.

From the following section on, we will introduce PVMark in detail.

\section{Adapting Watermark Schemes for PVMark}
Since the original watermark schemes contain procedures that are challenging to construct proofs, we adapt them to almost equivalent, ZKP-friendly forms. The adaptation preserves the original schemes' foundational properties of effectiveness, fidelity, and robustness (see experimental results in the supplementary file).

\subsection{Hash-Based Vocabulary Partitioning}
A key step in watermark embedding is to assign random ranks to candidate tokens in realizing the random partition of vocabulary into green and red lists (Step 2 in Table~\ref{tab:watermark_embed}). Green tokens are selected with a probability higher than random chance.

The original watermarking schemes primarily employ two methods for vocabulary partitioning. The first is random permutation, utilized in KGW~\cite{kirchenbauer2023watermark} and Segment-Watermark~\cite{qu2025provablyrobustmultibitwatermarking}: it randomly permutes all the candidate tokens in the vocabulary and marks the first $\gamma \cdot  |V|$ indexed tokens as the green list and the rest as the red list. Random permutation requires a mapping operation from random numbers to the range of the vocabulary size, which involves ZK-unfriendly modulo operations. 

The second is the tournament method, adopted by SynthID-Text~\cite{dathathri2024scalable}: it assigns random numbers to each token in the candidate token set by using a precomputed lookup table and a pseudo random function (PRF) instantiated by a linear congruential generator (LCG). Initially, a lookup table of size $2^{16}$ is randomly populated with $0$s and $1$s. Then, the LCG generates $\xi$ random numbers, which are mapped to the range of $[0, 2^{16} - 1]$ for retrieving values from the table. These values are then assigned to $g_1, ..., g_\xi$. Subsequently, all candidate tokens undergo multiple rounds of competition, where tokens with smaller $g$ values are eliminated iteratively. Tokens that remain in the final round are designated as green tokens. This process also involves ZK-unfriendly modulo operations in LCG.

To avoid modulo operations, we propose a Hash-based method for vocabulary partition. A straightforward and equivalent substitute for random permutation is random numbers sorting where we sort $\boldsymbol{r} = \{r_1, \ldots, r_{|V|} \}$ in numerical order, and take the tokens corresponding to the first $\gamma |V|$ indices in $\boldsymbol{r}$ to compose the green list. Unfortunately, the proof of sorting is still highly inefficient to implement. Hence we take an approximated approach by assigning hash-generated random numbers to candidate tokens. The vocabulary is divided into green/red lists by comparing these numbers with a fixed threshold $\gamma \cdot |\mathbb{F}|$, shown by the first row of Step 2 in Table~\ref{tab:watermark_embed}. Similarly, for the tournament method, we compare hash-generated random numbers with a fixed threshold $\frac{|\mathbb{F}|}{2}$, and map them to $0$ or $1$ depending on the comparison result (shown by the second row of Step 2 in Table~\ref{tab:watermark_embed}). However, our hash-based approach relies on the assumption that the hash-generated random numbers are uniformly distributed within the range, so that the fixed threshold could divide the vocabulary in proportion. We validate that the assumption mostly holds true in practice (see Sec.~\ref{sec_exp:randomness}).

In addition, we instantiate the PRF required for computing the random seed using a hash function. Similarly, all PRFs involved in the watermark detection are also instantiated via hash functions. When checking and counting green tokens, we determine whether the current token is a green token by comparing its corresponding random number with a fixed threshold, as shown in Step 2 of Table~\ref{tab:watermark_detection}.

\subsection{Selection of the Hash Function}
We select cryptographic hash functions as PRFs since they are high-quality pseudorandom number generators following the guidelines NIST SP 800-90A \cite{2811}, instead of non-cryptographic pseudorandom number generation algorithms such as Linear-feedback shift register \cite{tausworthe1965random}, Blum Blum Shub \cite{blum1986simple}, Mersenne Twister \cite{matsumoto1998mersenne}, Xorshift \cite{marsaglia2003xorshift}, etc. Although the non-cryptographic ones are faster by bit operations, they are not ZKP-friendly for incurring high costs in the verification of bitwise operations.

The most commonly used cryptographic hash functions include SHA256, Keccak256, BLAKE2, which take inputs of arbitrary length and map them to a fixed 256-bit output. These hash functions contain many bit operations which involve high ZKP-related costs. For instance, in verifying SHA256, the significant cost arises from its compression function that contains bit decompositions. Other hash functions are more ZK-friendly, including MiMC hash \cite{albrecht2016mimc}, Poseidon hash \cite{grassi2021poseidon} and Poseidon2 hash \cite{grassi2023poseidon2}. These hash functions are effective in finite fields applicable to ZKP and mostly consist of addition and multiplication operations which are relatively efficient to verify. For example, Poseidon hash is based on the sponge function, with a state composed of finite field elements and field operations including addition, multiplication and exponentiation, all of which can be easily converted into ZKP circuits.

To verify the uniformity and randomness of the cryptographic hash functions, we run the chi-square test to examine SHA256, Keccak256, BLAKE2, MiMC, Poseidon and Poseidon2. The results are presented in Sec.~\ref{sec_exp:randomness}. By tests, we found SHA256, Keccak256, or BLAKE2 are not applicable for \textit{PVMark} since the 256-bit value range of these hash does not match the finite field used by ZKP, and thus the hash values are hardly uniformly distributed on $\mathbb{F}$. MiMC, Poseidon and Poseidon2, hash functions that passed the test, are our final choices for PRFs.

\section{Building Blocks of PVMark}
\label{sec::building_blocks}
To prove watermark detection, it is required for \PVM~to generate proofs for the three steps in Table~\ref{tab:watermark_detection}. We observe that in Step 3 for computing the score/message, the only unknowns are \( y_G \), \( S_g \), and COUNT, while other quantities are public. Hence, as long as the verifier obtains the correct \( y_G \), \( S_g \), and COUNT, it can derive the score or watermark message to determine whether the given text contains the watermark or extract the embedded watermark. Therefore, \PVM~only needs to prove the correctness of \( y_G \), \( S_g \), and COUNT, i.e., generate correctness proofs for Step 1 and Step 2, yet without exposing \( sk \) or the green list $G$. Next, we will present in detail the arithmetization of the two steps by the example of PLONKish.

\subsection{Arithmetize Step 1 in Table~\ref{tab:watermark_detection}}
\label{sec:arithmetize_step_1}

Since we have instantiated the PRF by a hash function, we arithmetize the hash to prove Step 1 of Table~\ref{tab:watermark_detection}.

\textbf{Hash functions.}
The constraints $\mathcal{C}_{hash}(\text{Output}; \text{Inputs})$ differ per hash function. We take two common ZK-friendly hash functions, MiMC hash and Poseidon hash, as examples. The MiMC function hashing $X$ is defined as $H_\text{mimc}(X) = (f_{\phi-1} \circ f_{\phi-2}\circ \ldots \circ f_{0})(X) + k$
where $\phi$ and $k$ denote the number of rounds and the secret key in MiMC hash, respectively. The round function $f_i$ is defined as
$f_i(Y_i) = (Y_i+k+C_i)^{\text{Pow}}$
where $C_i$ is the round constant for round $i$ and $Y_0 = X$. To construct the constraints of $f_i(\cdot)$, we introduce a new power operation selector $s_\text{Pow}$ to rewrite Eq.~\eqref{eq:standard_plonkish} as
\begin{equation}\label{eq:pow_constraints}
	s_\text{Pow} \cdot (x_L + x_R + c)^{\text{Pow}} + s_O \cdot x_O = 0,
\end{equation}
and set $s_\text{Pow}=1, s_O=-1,x_L=Y_i,x_R=k,c=C_i$ and $x_O=f_i(Y_i)$.
We construct the MiMC hash by reusing Eq.~\eqref{eq:pow_constraints} $\phi$ times, and use copy constraints to ensure the consistency of input and output in function compositions. 

The construction of constraints for Poseidon hash is relatively simpler. Poseidon hash is composed of three components -- ARC($\cdot$), S-box($\cdot$), and Mix($\cdot$), for a single round. Specifically, ARC($\cdot$) adds a constant to all inputs, S-box($\cdot$) performs exponentiation wherein the last input of partial round $R_p$ and all inputs of the full round $R_f$ are raised to the power of 5. %
Similar to Eq.~\eqref{eq:pow_constraints}, we use $s_\text{Pow}$ selector to construct the S-box function. In Mix($\cdot$), a pre-calculated matrix is used to perform vector-matrix multiplication operations for all inputs. We also use copy constraints to ensure the consistency of input and output between each round.

\textbf{Token-position mapping.} In the multibit scheme, Step 1 requires looking up the position $p$ for $\boldsymbol{y}_{\psi}$ from the token-to-position table $M$. \PVM~provides a proof for the lookup as it is a part of the counting. However, such a mapping $M$ cannot be disclosed otherwise would allow adversary to infer about the relation between token and the value in $\boldsymbol{msg}[p]$ based on the publicly available COUNT matrix. This would tempt the adversary to remove the watermark from $\boldsymbol{y}$.

To check the correctness of a set of non-exposable token-position mappings in the watermark detection, we require the prover to precompute the hash values for all possible token-position mappings, use them as leaf nodes to construct a merkle tree, and publish the root value in advance. Thus the proof of a set of mappings reduces to the proof of membership of hash values in the merkle tree.

The membership proof requires the prover to recalculate the root value by the values of the sibling nodes in the path from the given leaf node to the merkle tree root. If the newly calculated root value matches the publicly disclosed one, the given leaf node must exist in the merkle tree. The specific constraints required for membership proof $\mathcal{C}_{mem}$ are as follows:

\begin{equation} \label{eq:c4}
	\mathcal{C}_{mem}(y, p, \text{ROOT})=
	\begin{cases}
		H_{0} = \mathcal{C}_{hash}(y, p, sk), \\
		path_i \cdot (1 - path_i) = 0, \\
		\begin{aligned}[t] &H_i = 
			path_i \cdot \mathcal{C}_{hash}(H_{i-1}, \text{Sib}_i) \\
			&+ (1 - path_i) \cdot \mathcal{C}_{hash}(\text{Sib}_i, H_{i-1}),
		\end{aligned} \\
		H_L - \text{ROOT} = 0,
	\end{cases}
\end{equation}
for $i = \{1, \ldots, L\}$ where $y, p, \text{ROOT}, sk$ denote the token index, position index, a publicly known Merkle tree root value, and the secret key, respectively. The $path_i \in \{0,1\}$ refers to the left/right child of the Merkle tree nodes, which influences the order of hash inputs during computation.  $Sib_i$ denotes the value of sibling node along the path from the given leaf node to the merkle tree root. %

\subsection{Arithmetize Step 2 in Table~\ref{tab:watermark_detection}}

In Step 2, we need to compute the corresponding random number for the current token \( y \), compare this random number with a fixed threshold, and perform an accumulation operation if the token is a green token. Hence we arithmetize the comparison and summation to prove the step.

\textbf{Comparison.} Assume that $X_1$ and $X_2$ are two numbers on finite field $\mathbb{F} = \mathbb{Z} \bmod p$ where $p$ is a large prime that $p > 2^N$. Without loss of generality, $X_1 < X_2$ if and only if $X_1 - X_2 \in [p - 2^{N-1}, p)$. Letting $X_3 = X_1 - X_2 - (p - 2^{N-1})$, we have $X_3 \in [0, 2^{N-1})$ and hence bit decomposition $\mathcal{C}_{bit}$ \cite{setty2012taking} can be used to check whether $X_3$ is in this range. The constraints for comparison $\mathcal{C}_{comp}(X_1 < X_2)$ can be written as follows:
\begin{equation}
	\label{eq:compare}
	\mathcal{C}_{comp}(X_1 < X_2)=
	\begin{cases}
		X_3 = X_1 - X_2 - (p - 2^{N-1}), \\
		\mathcal{C}_{bit}(X_3, N-1, \boldsymbol{b}_{X_3}),
	\end{cases}
\end{equation}
where $\boldsymbol{b}_{X_3} = [ b_0, b_1, \ldots, b_{N-1}]$ denotes the binary representation of $X_3$. %
$\mathcal{C}_{comp}(X_1 < X_2) = 0$ suggests $X_1 < X_2$ holds and all constraints included are satisfied.

To reduce the number of constraints and thus the overhead of ZKP, we optimize the proof of $\mathcal{C}_{bit}(\cdot)$ employing the lookup table, a technique that trades space for time. Specifically, traditional bit decomposition needs to check the correctness of the binary form bit by bit, i.e., using $b_i \cdot (1-b_i) = 0$ to check whether $b_i$ equals $0$ or $1$, followed by checking whether the weighted sum of $b_i$s equals $X_3$. To avoid the exorbitant overhead of bit checking, we propose to check $X_3$ by groups of 8 bits, i.e., verifying whether each group lies within the range of $[0, 255]$, and subsequently check if the weighted sum of all groups equals $X_3$. For each group, we predefine the range of $[0, 255]$ using the lookup table to reduce the number of constraints.

\textbf{Summation.} We take KGW as an example to illustrate the arithmetization method of summation. To count the number of green tokens $|\boldsymbol{y}|_{G}$, we introduce an auxiliary flag variable ${flg}_i \in \{0, 1\}$ to indicate whether token $y_i$ belongs to the green list $G_i$. The following constraint $\mathcal{C}_{flg}$ verifies the assignment of ${flg}_i$:
\begin{equation} \label{eq:c3}
	\mathcal{C}_{flg}({flg}_i, r_i,  \gamma |\mathbb{F}|)=
	\begin{cases}
		flg_i \cdot (1 - flg_i) = 0, \\
		flg_i \cdot \mathcal{C}_{comp}(r_i < \gamma |\mathbb{F}|) = 0, \\
		(1 - flg_i) \cdot \mathcal{C}_{comp}( \gamma |\mathbb{F}|< r_i) = 0,
	\end{cases}
\end{equation}
where $\gamma |\mathbb{F}|$ denotes the pre-selected threshold and $r_i$ represents the random number assigned to token $y_i$, which is computed by the aforementioned hash function and verified by $\mathcal{C}_{hash}$. Then we introduce a summation selector $s_\text{Sum}$ and rewrite Eq.~\eqref{eq:standard_plonkish} as
\begin{equation}\label{eq:sum_constraints}
	s_\text{Sum} \cdot (x_1 + x_2 + \ldots + x_n) + s_O \cdot x_O = 0,
\end{equation}
to construct the summation constraint $\mathcal{C}_{sum}(\text{Output}; \text{Inputs})$.

\section{Applications of PVMark}

In this section, we present how \PVM~applies to different watermarking schemes to gain public verifiability. The optimized version using recursive ZKP is also demonstrated with examples.

\begin{algorithm}[t]
	\renewcommand{\algorithmicrequire}{\textbf{Input:}}
	\renewcommand{\algorithmicensure}{\textbf{Output:}}
	\caption{Proof Construction for KGW / \underline{SynthID-Text}}
	\label{alg:zkp_sketch_for_zero_bit_detection}
	\begin{algorithmic}[1]
		\REQUIRE Secret key $sk$, watermarked text $\boldsymbol{y} = [y_0, \ldots, y_{n - 1}]$, recent context window width $\psi$, green list size $\gamma$, threshold $\gamma \cdot |\mathbb{F}|$ / \underline{$\frac{|\mathbb{F}|}{2}$}, the count result $|\boldsymbol{y}|_{G}$ / \ul{$S_g$}, \underline{the number of g-values assigned to each token $\xi$}. 
		\ENSURE Proof $\Pi$.
		
		\FOR{$i = \psi, \psi + 1, \ldots, n-1$}
		\STATE \COMMENT{Arithmetize Step 1}
		\STATE Build $\mathcal{C}_{hash}(sd_i; sk, \boldsymbol{y}_{\psi})$.
		\STATE \COMMENT{Arithmetize Step 2}
		\STATE Build $\mathcal{C}_{hash}(g^{y_i}; sd_i, y_i)$ /  \ul{$\mathcal{C}_{hash}(g_k^{y_i}; sd_i, y_i, k)$ for $k \in \{1, \ldots, \xi \}$}.
		\STATE Set $flg^i = \mathds{1}(g^{y_i} < \gamma \cdot |\mathbb{F}|)$ / \ul{$flg^i_k = \mathds{1}(g_k^{y_i} < \frac{|\mathbb{F}|}{2})$ for $k \in \{1, \ldots, \xi \}$}.
		\STATE Build $\mathcal{C}_{flg}(flg^i, g^{y_i}, \gamma \cdot |\mathbb{F}|)$ / \ul{$\mathcal{C}_{flg}(flg^i_k, g_k^{y_i}, \frac{|\mathbb{F}|}{2})$ for $k \in \{1, \ldots, \xi \}$}.
		\ENDFOR
		\STATE Build $\mathcal{C}_{sum}((|\boldsymbol{y}|_{G}$ / \ul{$S_g$}; $Flg)$ where $Flg = \{flg^{i} | i \in \{ \psi, \ldots, n - 1 \}\}$ / \ul{$\{flg_k^{i} | i \in \{ \psi, \ldots, n - 1 \}, k \in [\xi]\}$}.
		\STATE $\mathcal{D}$ invokes ZKP.Prove$(u, w, \boldsymbol{\mathcal{C}})$ to generate proof $\pi$ where $u = \{\boldsymbol{y}, \psi, \gamma |\mathbb{F}| / \underline{\frac{|\mathbb{F}|}{2}}, |\boldsymbol{y}_{G}| / \underline{S_g}, \underline{\xi} \}$ are the public input, $w = \{sk\}$ denotes the private input and $\boldsymbol{\mathcal{C}}$ denotes the set of all constraints.
		\STATE Return $\Pi = \{u, \pi\}$.
	\end{algorithmic}	
\end{algorithm}

\subsection{Verifying Zero-\& Multi-Bit Watermarks}
\PVM~is applied right after the watermark detection phase by prover $\mathcal{D}$, requiring the detection results $|\boldsymbol{y}|_{G}$, $S_g$, COUNT and other constants as inputs. The output of proof construction is a proof $\pi$ provided to verifier $\mathcal{V}$ who either passes or rejects the proof. Since the proving and verification follow the conventional ZKP protocols, we mainly focus on the proof construction. Hence the proof construction for KGW and SynthID-Text is introduced in Alg.~\ref{alg:zkp_sketch_for_zero_bit_detection} whereas that for Segment-Watermark is described in Alg.~\ref{alg:zkp_sketch_for_multi_bit_detection}.

We show the proof construction of KGW and SynthID-Text in Alg.~\ref{alg:zkp_sketch_for_zero_bit_detection} with distinctions underlined. The algorithm mainly includes building $\mathcal{C}_{hash}$, $\mathcal{C}_{flg}$, and $\mathcal{C}_{sum}$. Then $\mathcal{D}$ generates proof $\pi$ for all the constraints and releases $\pi$ and the public inputs to $\mathcal{V}$ to carry out proving and verification following ZKP protocols.

\textbf{Combining hash functions.} 
In the original KGW scheme, $\psi$ is set to 1, which allows us to make further adaptations by combining two hash functions (Line 3 and 5 of Alg.~\ref{alg:zkp_sketch_for_zero_bit_detection}) into a three-to-one hash, thereby saving ZKP costs. Specifically, in round $i$, the random number $g^{y_i}$ corresponding to the current token $y_i$ is computed using two hash functions with two arguments: one calculates a random seed $sd_i$ from $sk$ and the previous token's index $y_{i-1}$, and the other uses $sd_i$ and the current candidate token’s index $y_i$ to compute $g^{y_i}$ (i.e., Step 1 and 2 in Table~\ref{tab:watermark_detection}). We propose to replace them with a single function with three arguments:  
\[
g^{y_i} = \text{Hash}(sk, y_{i-1}, y_i),
\]  
which maps the tuple of the secret key, previous token index, and current candidate token’s index to a single hash value. Therefore, for KGW, Lines 3 and 5 in Alg.~\ref{alg:zkp_sketch_for_zero_bit_detection} can be merged into `Build $\mathcal{C}_{hash}(g^{y_i}; sk, y_{i-1}, y_i)$.' This optimization effectively reduces the circuit size and the ZKP costs in the detection process. 

Note that in the original SynthID-Text, $\psi$ is set to be greater than 1, e.g., $\psi = 4$. When using three-to-one hash functions to construct the constraint for Line 3 of Alg.~\ref{alg:zkp_sketch_for_zero_bit_detection}, we combine pairs of hash functions into one. Therefore, $\mathcal{C}_{hash}(sd_i; sk, \boldsymbol{y}_{\psi})$ is defined as the set  
\[
\left\{
\begin{aligned}
	&\mathcal{C}_{hash}(H_1^{(i)}; sk, \boldsymbol{y}_{\psi}^{(i)}[0], \boldsymbol{y}_{\psi}^{(i)}[1]), \\
	&\mathcal{C}_{hash}(H_2^{(i)}; H_1^{(i)}, \boldsymbol{y}_{\psi}^{(i)}[2], \boldsymbol{y}_{\psi}^{(i)}[3]), \\
	&\hspace{2cm} \vdots \\
	&\mathcal{C}_{hash}(sd_i; H_{\frac{\psi}{2} - 1}^{(i)}, \boldsymbol{y}_{\psi}^{(i)}[\psi - 2], \boldsymbol{y}_{\psi}^{(i)}[\psi - 1])
\end{aligned}
\right\}
\]  
where $H^{(i)}$ denotes the corresponding hash results.

\begin{algorithm}[t]
	\renewcommand{\algorithmicrequire}{\textbf{Input:}}
	\renewcommand{\algorithmicensure}{\textbf{Output:}}
	\caption{Proof Construction for Segment-Watermark}
	\label{alg:zkp_sketch_for_multi_bit_detection}
	\begin{algorithmic}[1]
		\REQUIRE Secret key $sk$, watermarked text $\boldsymbol{y} = [y_0, \ldots, y_{n - 1}]$, recent context window width $\psi$, green list size $\gamma$, threshold $\gamma \cdot |\mathbb{F}|$, token-to-position mapping $M: [0, |V| - 1] \rightarrow [0, \hat{n} - 1]$, the root value ROOT of the merkle tree corresponding to $M$, the count result matrix COUNT $\in \mathbb{Z}^{\hat{n} \times 2^{\hat{m}}}$.
		\ENSURE Proof $\Pi$.
		
		\FOR{$i = \psi, \psi + 1, \ldots, n-1$}
		\STATE Build $\mathcal{C}_{mem}(\boldsymbol{y}_{\psi}, p_i, \text{ROOT})$.
		\FOR{$j = 0, 1, \ldots, 2^{\hat{m}-1}$}
		\STATE Build $\mathcal{C}_{hash}(sd_i^j; sk, \boldsymbol{y}_{\psi}, j)$.
		\STATE Build $\mathcal{C}_{hash}(g_j^{y_i}; sd_i^j, y_i)$.
		\STATE Set $flg_j^i = \mathds{1}(g_j^{y_i} < \gamma \cdot |\mathbb{F}|)$.
		\STATE Build a constraint using an addition gate for $\text{COUNT}[p_i][j] \mathrel{+=} flg_j^{i}$.
		\ENDFOR
		\ENDFOR
		\STATE $\mathcal{D}$ invokes ZKP.Prove$(u, w, \boldsymbol{\mathcal{C}})$ to generate proof $\pi$ where $u = \{\boldsymbol{y}, \psi, \gamma |\mathbb{F}|, \text{ROOT}, \text{COUNT} \}$ are the public input, $w = \{sk, M\}$ denotes the private input and $\boldsymbol{\mathcal{C}}$ denotes the set of all constraints.
		\STATE Return $\Pi = \{u, \pi\}$.
	\end{algorithmic}	
\end{algorithm}

We also present in Alg.~\ref{alg:zkp_sketch_for_multi_bit_detection} the proof construction of Segment-Watermark, which has an additional contraint $\mathcal{C}_{mem}$ for checking the token-to-position mapping, compared with Alg.~\ref{alg:zkp_sketch_for_zero_bit_detection}.

\subsection{Optimization with Recursive ZKP}
\begin{algorithm}
	\renewcommand{\algorithmicrequire}{\textbf{Input:}}
	\renewcommand{\algorithmicensure}{\textbf{Output:}}
	\caption{Proof Construction w/ Recursive ZKP for KGW}
	\label{alg:recursive}
	\begin{algorithmic}[1]
		\REQUIRE Same inputs with Alg.~\ref{alg:zkp_sketch_for_zero_bit_detection}, number of tokens per instance $N_t$, number of foldings $N_f$ and the secret $s_H$.
		\ENSURE proof $\Pi$.
		
		\STATE Initialize the number of green tokens $|\boldsymbol{y}|^{(1)}_{G}=0$ and $H^{(1)}= \text{Hash}(0, s_H)$.
		\FOR{$i = 1, 2, \ldots, N_f$}
		\STATE Build $\mathcal{C}_{hash}(|\boldsymbol{y}|^{(i)}_{G}; s_H, H^{(i)} )$ for consistency check.
		\FOR{$j = 1, 2, \ldots, N_t$}
		\STATE Invoke Lines 2 to 7 in Alg.~\ref{alg:zkp_sketch_for_zero_bit_detection} to build the constraints $\mathcal{C}_{hash}$ and $\mathcal{C}_{flg}$.
		\ENDFOR
		\STATE Accumulate the increment of $|\boldsymbol{y}|_{G}^{(i)}$ to $|\boldsymbol{y}|_{G}^{(i+1)}$ by adding up $flg^j$ for $1 \leq j \leq N_t$ and establish the addition constraints.
		\STATE Build $\mathcal{C}_{hash}(|\boldsymbol{y}|^{(i+1)}_{G}; s_H, H^{(i+1)} )$ for consistency check where $H^{(i+1)} = \text{Hash}(|\boldsymbol{y}|^{(i+1)}_{G}, s_H)$.
		\STATE Collect all the above constraints, represent them using relaxed R1CS, and derive a relaxed R1CS instance $Ins_{(i)}$.
		\IF{$i \neq 1$}
		\STATE Fold $Ins_{(i)}$ with $Ins_{(i-1)}^{\prime}$ by establishing the constraints for folding process as in Eq.~\eqref{eq:fold1}, ~\eqref{eq:fold2} and ~\eqref{eq:fold3} to obtain a folded instance $Ins_{(i)}^{\prime}$ ($Ins_{(1)}^{\prime} = Ins_{(1)}$). 
		\ENDIF
		\ENDFOR
		\STATE Obtain the final instance $Ins_{(N_f)}^{\prime}$.
		\STATE Build $\mathcal{C}_{hash}(|\boldsymbol{y}|^{(N_f +1)}_{G}; s_H, H^{(N_f+1)} )$ and obtain an additional instance $Ins_{ad}$, where $|\boldsymbol{y}|^{(N_f +1)}_{G}$ and $H^{(N_f+1)} $ are public inputs, and $s_H$ is private input.
		\STATE $\mathcal{D}$ invokes ZKP.Prove to generate proof $\pi_1$ and $\pi_2$ for $Ins_{(N_f)}^{\prime}$ and $Ins_{ad}$, respectively.
		\STATE Return $\Pi = \{\pi_1, \pi_2\}$.
	\end{algorithmic}	
\end{algorithm}

We observe that for a chosen watermarking scheme and a selected hash function, the process of detecting the watermark for any token in text $\boldsymbol{y} = [y_0, \ldots, y_{n - 1}]$ follows the same steps, as shown in Line 2 to 7 of Alg.~\ref{alg:zkp_sketch_for_zero_bit_detection} and Line 2 to 8 of Alg.~\ref{alg:zkp_sketch_for_multi_bit_detection}. This observation naturally leads us to construct the detection process by incrementally verifiable computation (IVC), as shown in Eq.~\eqref{eq:ivc_form}. Take the KGW as an example, we reformulate the calculation of the sum of green tokens as
\begin{equation}
	\label{eq:f_1}
	|\boldsymbol{y}|_{G}^{(i + 1)} = F(|\boldsymbol{y}|_{G}^{(i)}, y_i).
\end{equation}
Within $F$, we compute the random number corresponding to $y_i$ and compare it with the pre-selected threshold $\gamma |\mathbb{F}|$. If the random number is less than the threshold, $y_i$ is a green token, and we set $|\boldsymbol{y}|_{G}^{(i+1)} = |\boldsymbol{y}|_{G}^{(i)} + 1$. After $n - 1$ iterations, $|\boldsymbol{y}|_{G}^{(n)}$ represents the count of green tokens in text $\boldsymbol{y}$.

Given the incremental calculation, we can efficiently generate zero-knowledge proofs using Nova's folding scheme, turning the detection of each token in $\boldsymbol{y}$ as a relaxed R1CS instance. %
These instances are fold into one final instance by Eq.~\eqref{eq:fold1}, ~\eqref{eq:fold2} and ~\eqref{eq:fold3}. Finally, it suffices to use Nova to prove the correctness of the final instance and each folding process, which are represented as relaxed R1CS instances. However, a challenge arises as Nova requires the incremental value $|\boldsymbol{y}|_{G}^{(i)}$ to be public, which does not meet the watermarking requirement that anything about the green token list should be private to prevent removal or ambiguity attacks.

To address this issue, we propose to disclose the hash of $|\boldsymbol{y}|_{G}^{(i)}$ instead of the value. This modifies Eq.~\eqref{eq:f_1} to
\begin{equation}
	\label{eq:f_2}
	\text{Hash}(|\boldsymbol{y}|_{G}^{(i + 1)}, s_H) = F(\text{Hash}(|\boldsymbol{y}|_{G}^{(i)}, s_H), y_i),
\end{equation}
where $s_H$ is a private input introduced to prevent brute-force attacks on the hash. Since hashing is added, each instance adds two consistency check to verify 1) whether the private input $|\boldsymbol{y}|_{G}^{(i)}$ of this round agrees with the output from the previous round, i.e., Line 3 of Alg.~\ref{alg:recursive}; 2) whether the public hash value $\text{Hash}(|\boldsymbol{y}|_{G}^{(i+1)}, s_H)$ truely hashes the output of this round ($|\boldsymbol{y}|_{G}^{(i+1)}$), i.e., Line 8 of Alg.~\ref{alg:recursive}. 

Obtaining $\text{Hash}(|\boldsymbol{y}|_{G}^{(n)}, s_H)$ through $n-1$ iterations of folding, we construct an additional instance on the hash value $\text{Hash}(|\boldsymbol{y}|_{G}^{(n)}, s_H)$ and the public input $|\boldsymbol{y}|_{G}^{(n)}$, with $s_H$ being the private input, to prove the consistency between $\text{Hash}(|\boldsymbol{y}|_{G}^{(n)}, s_H)$ and the public $|\boldsymbol{y}|_{G}^{(n)}$, shown in Line 15 of Alg.~\ref{alg:recursive}. The value of $|\boldsymbol{y}|_{G}^{(n)}$ is used for subsequent watermark score calculation.

In practical applications, there is always a tradeoff between the size of each instance and the number of instances, which leaves design choices to make. Overall, for the same computation, the larger the instance size, the higher the cost per instance but there would be fewer instances to verify. In \PVM, the choice is the number of tokens to be verified per instance. Intuitively, the folding scheme can save ZKP cost if the cost of proving an instance exceeds the cost of proving a folding. But the savings differ on the choices. Hence we analyze the tradeoff per case, as shown in Sec.~\ref{sec:efficiency}. 

We sum up our recursive ZKP for the detection of KGW in Alg.~\ref{alg:recursive}. The recursive ZKP for the detection of SynthID-Text is similar and presented in Alg.4 in the supplementary file.

Note that in the multi-bit watermarking scheme, the detection processes for different positions are not IVC processes, since the final value of the count result at one position is not equal to the initial value of the count result of the next position. In contrast, the detection process for the same position follows the paradigm of IVC. Therefore, we require the prover to group all tokens in the given text $\boldsymbol{y}$ according to their corresponding positions before generating the proof, and then use recursive ZKP to generate correctness proofs for the detection processes of each position in parallel. The recursive ZKP for the detection of Segment-Watermark is presented in Alg.5 in the supplementary file.

\subsection{Security Analysis}
We claim that Alg.~\ref{alg:zkp_sketch_for_zero_bit_detection}, \ref{alg:zkp_sketch_for_multi_bit_detection}, \ref{alg:recursive}, and Alg.4 and 5 in the supplementary file all generate correctness proofs for the watermark detection process without exposing the $sk$, and the probability of using a wrong $sk^{\ast}$ to generate a proof that can deceive the verifier is negligible. This is guaranteed by the completeness, soundness and zero-knowledge property of zero-knowledge arguments, which have been proved in previous work~\cite{groth2016size, PLONK,kothapalli2022nova}. We provide the security proof for Alg.~\ref{alg:zkp_sketch_for_zero_bit_detection} in the supplementary file, and the proofs for the other four algorithms are similar. Moreover, the only viable attack, namely owners denying harmful text generation by using an inconsistent key, can be prevented by the key commitment in the Setup algorithm, where the commitment is implemented in practice using a cryptographic hash function.

\section{Evaluations}
Our adaption of the original watermarking schemes is based on the assumption that the hash function returns sufficiently random and uniformly distributed random values. Hence, for \PVM~to work in practice, we need to validate the assumption first. Besides, all properties of watermarking should be examined. Therefore, we aim to answer the following research questions by experiments on \textit{PVMark}: 1) Does our adaptation using hash functions as PRF satisfy the randomness and uniformity assumption? 2) Does \textit{PVMark} compromise the effectiveness, fidelity, and robustness of the original schemes (see the supplementary file)? 3) How efficient is \textit{PVMark}?

\begin{table}[tb]
	\centering
	\caption{All variants of the implemented watermark detection algorithm, where \ding{172}\ding{173}\ding{174}\ding{175} denotes Groth16, PlonK, halo2 and Nova, respectively.}
	\label{table:implementation}
	\resizebox{\linewidth}{!}{
		\begin{tabular}{ccccc}
			\toprule
			\textbf{Scheme} & \textbf{Hash variants} & \textbf{MiMC} & \textbf{Poseidon} & \textbf{Poseidon2} \\
			\midrule
			\multirow{2}{*}{\textbf{KGW}} & \textbf{Two-to-one} & \ding{172}\ding{173}\ding{174}\ding{175} & \ding{172}\ding{173}\ding{174}\ding{175} & \ding{172}\ding{173}\ding{175} \\
			& \textbf{Three-to-one} & \ding{172}\ding{173}\ding{174} & \ding{172}\ding{173}\ding{174} & / \\
			\textbf{SynthID-Text} & \textbf{Two-to-one} & \ding{172}\ding{173}\ding{175} & \ding{172}\ding{173}\ding{175} & \ding{172}\ding{173}\ding{175} \\
			\textbf{Segment-Watermark} & \textbf{Two-to-one} & \ding{175} & \ding{175} & \ding{175} \\
			\bottomrule
		\end{tabular}
	}
		\vspace{-0.4cm}	
\end{table}

\begin{table*}[htb]
	\centering
	\caption{The avalanche effect coefficient and the chi-square statistics of hash functions. The latter three are qualified.}
	\label{table:hash_random}
	\resizebox{\linewidth}{!}{
		\begin{tabular}{ccccccc}
			\toprule
			& \textbf{SHA256 hash} & \textbf{Keccak256 hash} & \textbf{BLAKE2 hash} & \textbf{MiMC hash} & \textbf{Poseidon hash} & \textbf{Poseidon2 hash} \\
			\midrule
			\textbf{Avalanche Effect Coeff.} & 0.498968 & 0.498948 & 0.498960 & 0.499570 & 0.499604 & 0.497579 \\
			\textbf{Percent Rate (\%)} & 0.206 & 0.210 & 0.208 & 0.086 & 0.079 & 0.484 \\
			\midrule
			\textbf{$\chi^2 \pm s$} & 370.322 $\pm$ 40.496 & 372.057 $\pm$ 40.068 & 370.755 $\pm$ 40.157 & 10.309 $\pm$ 5.321 & 10.218 $\pm$ 5.180 & 10.209 $\pm$ 5.175 \\
			\textbf{Success Rate (\%)} & 0 & 0 & 0 & 99.1 & 99.2 & 99.4 \\
			\bottomrule
		\end{tabular}
	}
	\vspace{-0.2cm}	
\end{table*}

\subsection{Implementation Details and Setup}

For the uniformity testing of hash functions, we implement SHA256, Keccak256, BLAKE2, MiMC, Poseidon, and Poseidon2 in Rust.

For the proof construction of watermark detection, we use circom \cite{circom}, a domain-specific language for defining arithmetic circuits, to construct constraints based on {\tt circomlib} \cite{circomlib}. These constraints can be compiled into R1CS and PLONKish circuits through compiler module in circom. ZKP protocol Groth16 is used to verify R1CS whereas PLONK is used to verify PLONKish. We also implemented a variant of Plonk --- halo2 \cite{halo2}, the KZG-based version, which supports custom gates and lookup operations thereby having a smaller circuit and better efficiency. We implement MiMC and Poseidon hash chips for halo2 to build PLONKish circuits. For the implementation of recursive ZKP, we use {\tt Nova-Scotia} \cite{nova-scotia}, which can compile circom circuits to Nova \cite{nova}. All variants of the implemented watermark detection are shown in Table~\ref{table:implementation}.

For all ZKP protocols, we choose BN254, a pairing-friendly elliptic curve defined over a prime field that currently provides approximately 100-bit security level \cite{kim2016extended}, widely used in blockchain project such as Ethereum \cite{wood2014ethereum} and Algorand \cite{chen2017algorand}. BN254 offers faster computation due to its smaller field size compared to curves like BLS12–381. 

\textbf{Parameter settings.} For watermark embedding and detection, we follow the settings of the original work \cite{kirchenbauer2023watermark, dathathri2024scalable, qu2025provablyrobustmultibitwatermarking}. For KGW, we set the green list size $\gamma = 0.25$ and the recent context window width $\psi = 1$. For SynthID-Text, we set the recent context window width $\psi = 4$ and the number of g-values assigned to each candidate token $\xi = 30$. For Segment-Watermark, we set the green list size $\gamma = 0.5$, the recent context window width $\psi = 1$, the total number of positions $\hat{n}=6$ in $\boldsymbol{msg}$ and the bit length $\hat{m}=4$ per position.

All experiments are conducted on a Ubuntu 20.04 server with Intel(R) Xeon(R) Gold 6240C CPU with 256GB RAM and an NVIDIA GeForce RTX 3090 GPU.

\begin{figure*}[htb]
	\centering
	\includegraphics[width=0.99\linewidth]{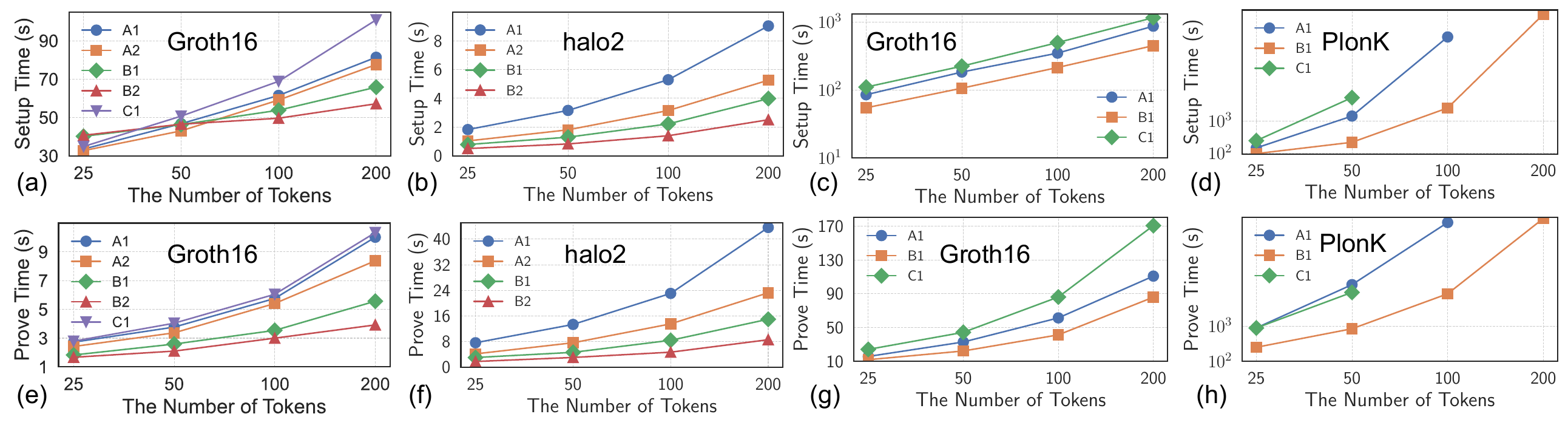}
	\caption{ZKP costs for \textit{PVMark} v.s. different numbers of tokens:  the left two columns are for detection by KGW and the right two columns are for detection by SynthID-Text. $A,B,C$ denote MiMC, Poseidon, Poseidon2 variants, respectively and $1$ and $2$ represent the use of two-to-one hash and three-to-one hash.}
	\label{fig::zkp_for_alg1}
\end{figure*}

\begin{figure*}[htb]
	\centering
	\includegraphics[width=0.99\linewidth]{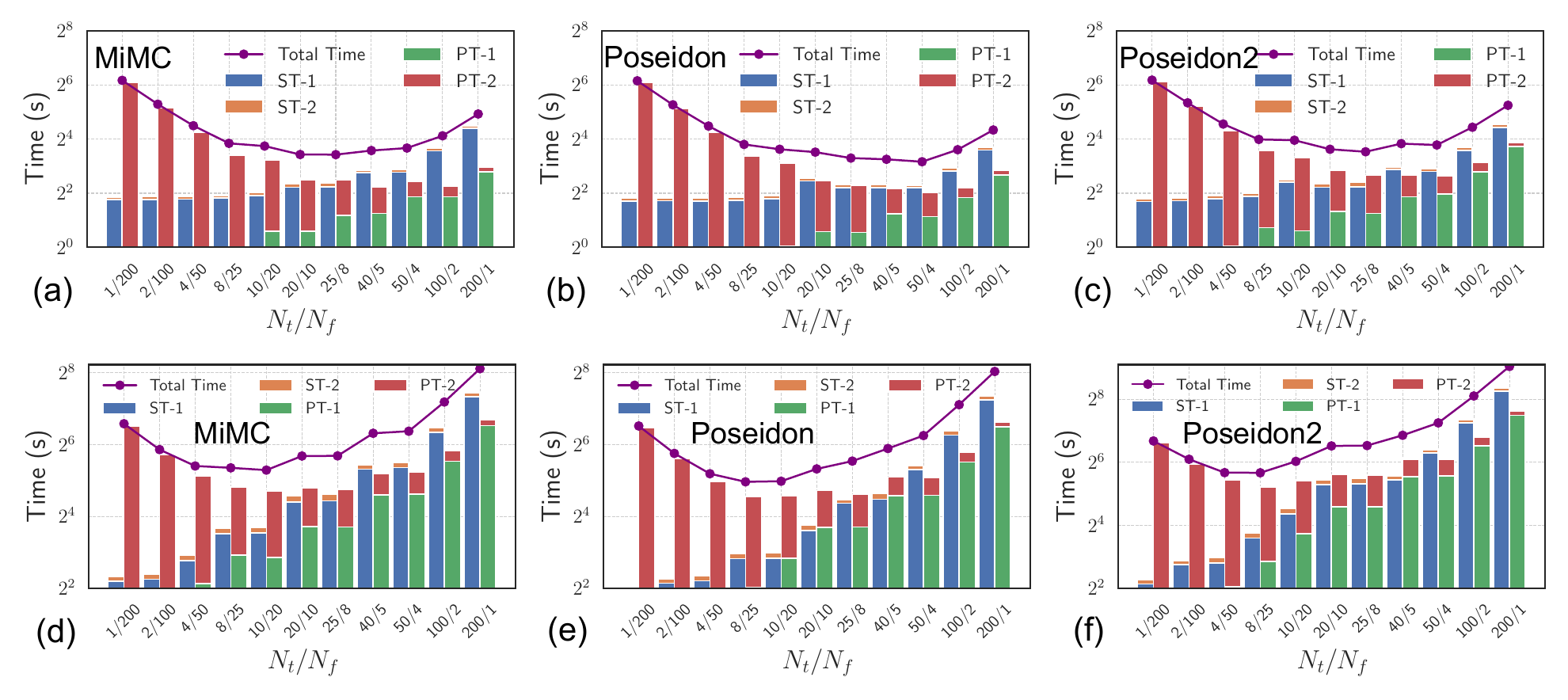}
	\caption{ZKP costs for Nova version of \textit{PVMark} v.s. varying  $N_t, N_f$ ($N_t \times N_f = 200$ tokens): the upper row is for detection by KGW and the lower row is for detection by SynthID-Text.  ST-1, ST-2 denote the setup time to prove the final instance and the folding process, respectively, and PT-1, PT-2 are their prove time correspondingly. The total time cost is ST-1 + ST-2 + PT-1 + PT-2. }
	\label{fig::nova_for_alg2}
		\vspace{-0.5cm}
\end{figure*}

\begin{figure*}[htb]
	\centering
	\includegraphics[width=0.99\linewidth]{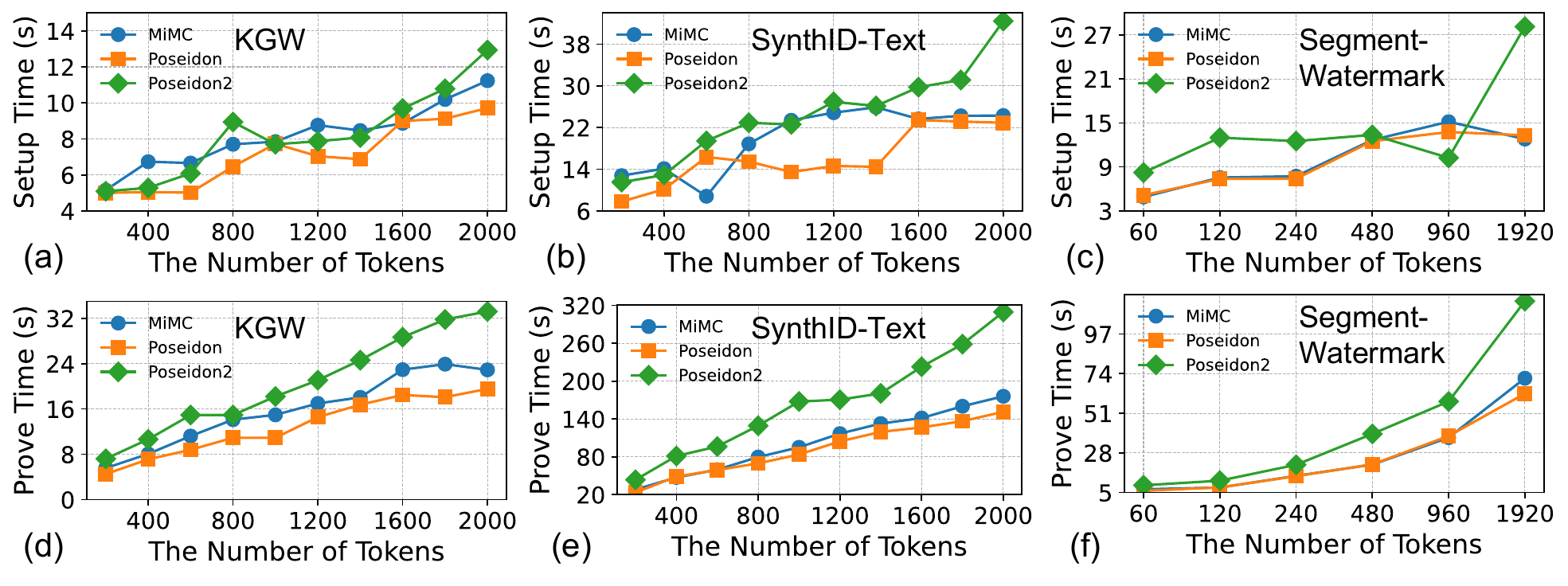}
	\caption{ZKP costs for Nova version of \textit{PVMark} v.s.  different numbers of tokens: the setting of \( N_t \) is such that the sum of setup time and prove time is minimized.}
	\label{fig::nova_for_large}
		\vspace{-0.3cm}
\end{figure*}

\begin{table*}[htbp]
	\centering
	\caption{Optimal \( N_t \) of \textit{PVMark} with Nova version v.s. different numbers of tokens (in the format of MiMC/Poseidon/Poseidon2 variants)}
	\label{tab:optimal_para}
	\resizebox{0.95\linewidth}{!}{
		\begin{tabular}{lcccccccccc}
			\toprule
			\multirow{2}{*}{\textbf{Scheme}} & \multicolumn{10}{c}{\textbf{Number of tokens  for watermark detection and verification}} \\
			\cline{2-11} 
			& 200 & 400 & 600 & 800 & 1000 & 1200 & 1400 & 1600 & 1800 & 2000 \\
			\midrule
			\textbf{KGW}          & 25/40/20 & 40/40/25 & 40/50/30 & 40/50/40 & 50/100/50 & 60/75/48 & 70/70/50 & 50/80/50 & 72/90/60 & 80/100/80 \\
			\textbf{SynthID-Text} & 8/8/5    & 10/8/5   & 12/15/8  & 16/16/10 & 20/20/10  & 20/20/12 & 20/20/14 & 20/25/16 & 25/25/18 & 25/25/20 \\
			\bottomrule
		\end{tabular}
	}
	\vspace{-0.4cm}	
\end{table*}

\begin{table}[htbp]
	\centering
	\caption{Optimal \( N_t \) of \textit{PVMark} with Nova version v.s. different numbers of tokens (in the format of MiMC/Poseidon/Poseidon2 variants)}
	\label{tab:segment-watermark-nt}
	\resizebox{0.98\linewidth}{!}{
	\begin{tabular}{lcccccc}
		\toprule
		\multirow{2}{*}{\textbf{Scheme}} & \multicolumn{6}{c}{\textbf{Number of tokens}} \\
		\cline{2-7} 
		& 60 & 120 & 240 & 480 & 960 & 1920 \\
		\midrule
		\textbf{Segment-Watermark} & 1/1/1 & 2/2/2 & 2/2/2 & 4/4/2 & 5/5/1 & 5/8/5 \\
		\bottomrule
	\end{tabular}
}
\vspace{-0.4cm}	
\end{table}

\subsection{Randomness and Uniformity of Hashes}
\label{sec_exp:randomness}
We first evaluate the randomness and uniformity of hash functions to validate the proposal of using hash functions as pseudorandom number generators and the rationality of using a pre-selected threshold to divide the green and red tokens, or assign g-values.

We use the avalanche effect coefficient and chi-square test to evaluate the randomness and uniformity of hash functions. The avalanche effect is a desirable property of cryptographic hash functions, suggesting an average of one-half of the output bits should change whenever a single input bit is flipped \cite{webster1985design}. A coefficient closer to 0.5 indicates better adherence to the avalanche effect property. We compute the avalanche effect coefficient by \cite{Avalanche_Alg}. Let $N_1, N_2$ denote the number of input bits and output bits, respectively. We set $N_1 = N_2 = 256$ for SHA256, Keccak256, BLAKE2 hash and $N_1 = N_2 = 254$ for MiMC, Poseidon, Poseidon2 hash, and execute the algorithm $5,000,000$ times to take the average. We show the closeness between the coefficient $c$ and $0.5$ by calculating the percentage of the difference, i.e., $\frac{|c - 0.5|}{0.5}$. A smaller percentage indicates better randomness.

To verify the uniformity, we conduct a chi-square goodness of fit test, a statistical hypothesis test. The null hypothesis is `all random numbers used for vocabulary partitioning in each round are uniformly distributed in the finite field $\mathbb{F}$ defined by BN254.' If the result $\chi^2$ is less than the threshold, the null hypothesis is accepted, meaning that the random numbers used for vocabulary division are uniformly distributed within the value range. We average the chi-square test statistics $\chi^2$ over $100,000$ iterations. In each iteration, we randomly select a secret key $sk \in \mathbb{F}$, a previous token index $y^{\prime} \in [0, |V| = 50265)$, and the current token index $\boldsymbol{y}^{\prime \prime} = [0, |V|)$ to generate $|V|$ random numbers. The statistic threshold is $26.296$ for significance level $\alpha = 0.05$.

The avalanche effect coefficient of hash functions and the percent rate of the coefficient v.s. $0.5$ are shown in the top part of Table~\ref{table:hash_random}. The average and standard deviation of multiple chi-squared tests statistic $\chi^2$, along with the success rate of passing the tests, are shown at the bottom of Table~\ref{table:hash_random}. It can be observed that the avalanche effect coefficients of the six types of hash functions are all close to 0.5, indicating sufficient randomness. On $\mathbb{F}$, MiMC, Poseidon, and Poseidon2 hash exhibit uniformity, thereby qualifying for our scheme.

Since we have validated the randomness and uniformity of hash functions used in \textit{PVMark}, our adaption theoretically does not affect the original watermarking scheme, i.e., an almost equivalent performance should be achieved with \textit{PVMark}. Due to space constraint, the effectiveness, fidelity and robustness results of the overall watermarking scheme are moved to the supplementary file. The conclusion is that the effectiveness, fidelity and robustness of watermarking is almost the same with or without \textit{PVMark}, which leaves the efficiency of \textit{PVMark} to be questioned.

\subsection{Efficiency}
\label{sec:efficiency}

\textbf{Proof generation for zero-bit watermarking schemes.}
Fig.~\ref{fig::zkp_for_alg1} and Fig.~\ref{fig::nova_for_alg2} present the ZKP costs for various \textit{PVMark} variants using four different ZKP protocols: Groth16, halo2, Plonk and Nova. ZKP costs include the setup time, prove time, verification time, and proof size. We do not present the verification time, as the slowest (PlonK) verification time does not exceed two seconds, which is far less than the setup and prove time.  

In Fig.~\ref{fig::zkp_for_alg1}, it is found that although the setup time of Groth16 is higher than that of halo2, their prove time shows the opposite. But still, the total time of halo2 is 1-5$\times$ faster than Groth16, mostly due to our careful design of circuit chips and the lookup table employed. PlonK, unfortunately, delivers exorbitant setup time and prove time, as its constraints are compiled by the universal Circom compiler, and restricted by the high-level programming language circom, preventing us from tailoring the circuit to the application. Among the variants of hash function, the three-to-one hash effectively reduces the cost compared to the two-to-one hash, and the saving grows with more tokens to be verified. Among all hash variants, Poseidon hash has the lowest overhead, exhibiting its suitablity for ZKP protocols. Compared to KGW, the detection of SynthID-Text is more costly. This is attributed to the more complicated process of Tournament sampling and thereby larger ZKP circuits. In fact, the circuit is so large that the memory demand is too high for us to fully measure the overhead of PlonK.

Since the original settings of KGW and SynthID-Text mostly involve a large number of tokens in watermark detection, we thus focus on the case of verifying $200$ or more tokens. In that case, recursive ZKP shows a much lower cost than others as in Fig.~\ref{fig::nova_for_alg2}. It is observed that the setup time for the folding process is almost constant and negligible compared to other parts. The setup time and prove time for the final instance increase with a rise in $N_t$ as a single instance grows larger. Meanwhile, the decrease in $N_f$ leads to a reduction in the prove time for the folding process. We find that for the MiMC, Poseidon, and Poseidon2 variants of \textit{PVMark}, setting $N_t$ to 20, 50, and 25 accordingly results in the minimum total overhead, which are around $2^4, 2^5$s for KGW, SynthID-Text, respectively. The overhead of detecting a larger number of tokens is shown in Fig.~\ref{fig::nova_for_large}. When detecting 2000 tokens, the least total time cost (setup and prove) is approximately $30$s and $175$s for KGW and SynthID-Text, respectively, where the optimal \(N_t\) is shown in Table~\ref{tab:optimal_para}. Hence we consider the running time is reasonable in a watermark verification setting.

\begin{table}[htb]
	\centering
	\caption{Proof sizes (KB) of the detection of KGW for 200 tokens. The most cost-effective $N_t, N_f$ are adopted for Nova.}
	\label{table:proof_size}
	\resizebox{0.85\linewidth}{!}{
		\begin{tabular}{ccccccc}
			\toprule
			\multirow{2}{*}{\textbf{ZKP}} & \multicolumn{2}{c}{\textbf{MiMC}} & \multicolumn{2}{c}{\textbf{Poseidon}} & \textbf{Poseidon2} \\
			\cmidrule(lr){2-3} \cmidrule(lr){4-5} \cmidrule(lr){6-6} 
			&  \textbf{2to1} & \textbf{3to1} &  \textbf{2to1} & \textbf{3to1} & \textbf{2to1} \\
			\midrule
			\textbf{Groth16} & 0.8 & 0.8 & 0.8  & 0.8  & 0.8  \\
			\textbf{halo2} & 14 & 14  & 14  & 14  & / \\
			\textbf{PlonK} & 2.3 & 2.3  & 2.3 &2.3 & 2.3  \\
			\textbf{Nova} & 10.2 & / & 10.5 & / & 10.5  \\
			\bottomrule         
		\end{tabular}
	}
	\vspace{-0.5cm}	
\end{table}

Table~\ref{table:proof_size} shows the proof size required in detection. Groth16 has the smallest proof size, less than 1KB. Actually, the proof sizes of Groth16, halo2, and PlonK are constant for 25/50/100/200 tokens. This is because Groth16's proof requires only 3 pairing checks, involving 3 group elements, regardless of the number of constraints. Both halo2 and PlonK use KZG commitments, of which the proof contains a limited group element, independent of the number of constraints. Nova utilizes Spartan \cite{setty2020spartan} at its core, a ZKP where proof size is logarithmically related to the number of constraints, but still the proof size is adequately small for 200 tokens.

\textbf{Proof generation for multi-bit watermarking schemes.}
Compared to the zero-bit watermark, it takes more constraints to verify the detection of Segment-Watermark. Hence we choose the Nova variant of \textit{PVMark} to generate proofs for cost consideration. It is worth noting that the detection for different positions of the message in Segment-Watermark can be performed in parallel. Thus we display in the rightmost of Fig.~\ref{fig::nova_for_large} the ZKP costs required for different numbers of tokens in verifying a single position of the watermark.

It can be observed that when the Poseidon hash function is used as the pseudorandom generator, the ZKP costs are minimized. When the number of tokens to be verified is $1920$ (i.e., for 6 positions, with each position requiring checking 320 tokens), the optimal setup time and prove time are approximately $13$ and $62$ seconds, which we consider to be a fully acceptable cost. The optimal \(N_t\) is shown in Table~\ref{tab:segment-watermark-nt}.

\section{Conclusion}
To resolve the trust issue in LLM watermark detection, we propose \textit{PVMark}, a plugin that enables zero-bit and multi-bit LLM watermarking schemes to gain public verifiability without compromising the original schemes' performance. The design is featured by the application of zero-knowledge proof to prove the correctness of watermark detection process, without revealing any credential. To achieve this, we redesigned the watermark embedding and detection schemes to align with ZKP, including introducing hash functions as PRFs, customizing the circuits, applying recursive ZKP, etc. We hope \textit{PVMark} not only contributes to the watermarking community, but also to a wider area of applied ZKP research.

\bibliographystyle{IEEEtran}
\bibliography{main}

\begin{thebibliography}{10}
\providecommand{\url}[1]{#1}
\csname url@samestyle\endcsname
\providecommand{\newblock}{\relax}
\providecommand{\bibinfo}[2]{#2}
\providecommand{\BIBentrySTDinterwordspacing}{\spaceskip=0pt\relax}
\providecommand{\BIBentryALTinterwordstretchfactor}{4}
\providecommand{\BIBentryALTinterwordspacing}{\spaceskip=\fontdimen2\font plus
\BIBentryALTinterwordstretchfactor\fontdimen3\font minus
  \fontdimen4\font\relax}
\providecommand{\BIBforeignlanguage}[2]{{%
\expandafter\ifx\csname l@#1\endcsname\relax
\typeout{** WARNING: IEEEtran.bst: No hyphenation pattern has been}%
\typeout{** loaded for the language `#1'. Using the pattern for}%
\typeout{** the default language instead.}%
\else
\language=\csname l@#1\endcsname
\fi
#2}}
\providecommand{\BIBdecl}{\relax}
\BIBdecl

\bibitem{pan2023riskmisinformationpollutionlarge}
\BIBentryALTinterwordspacing
Y.~Pan, L.~Pan, W.~Chen, P.~Nakov, M.-Y. Kan, and W.~Y. Wang, ``On the risk of
  misinformation pollution with large language models,'' 2023. [Online].
  Available: \url{https://arxiv.org/abs/2305.13661}
\BIBentrySTDinterwordspacing

\bibitem{vasilatos2023howkgptinvestigatingdetectionchatgptgenerated}
\BIBentryALTinterwordspacing
C.~Vasilatos, M.~Alam, T.~Rahwan, Y.~Zaki, and M.~Maniatakos, ``Howkgpt:
  Investigating the detection of chatgpt-generated university student homework
  through context-aware perplexity analysis,'' 2023. [Online]. Available:
  \url{https://arxiv.org/abs/2305.18226}
\BIBentrySTDinterwordspacing

\bibitem{kirchenbauer2023watermark}
J.~Kirchenbauer, J.~Geiping, Y.~Wen, J.~Katz, I.~Miers, and T.~Goldstein, ``A
  watermark for large language models,'' in \emph{International Conference on
  Machine Learning}.\hskip 1em plus 0.5em minus 0.4em\relax PMLR, 2023, pp.
  17\,061--17\,084.

\bibitem{dathathri2024scalable}
S.~Dathathri, A.~See, S.~Ghaisas, P.-S. Huang, R.~McAdam, J.~Welbl, V.~Bachani,
  A.~Kaskasoli, R.~Stanforth, T.~Matejovicova \emph{et~al.}, ``Scalable
  watermarking for identifying large language model outputs,'' \emph{Nature},
  vol. 634, no. 8035, pp. 818--823, 2024.

\bibitem{lee2024wrotecodewatermarkingcode}
\BIBentryALTinterwordspacing
T.~Lee, S.~Hong, J.~Ahn, I.~Hong, H.~Lee, S.~Yun, J.~Shin, and G.~Kim, ``Who
  wrote this code? watermarking for code generation,'' 2024. [Online].
  Available: \url{https://arxiv.org/abs/2305.15060}
\BIBentrySTDinterwordspacing

\bibitem{wang2024codablewatermarkinginjectingmultibits}
\BIBentryALTinterwordspacing
L.~Wang, W.~Yang, D.~Chen, H.~Zhou, Y.~Lin, F.~Meng, J.~Zhou, and X.~Sun,
  ``Towards codable watermarking for injecting multi-bits information to
  llms,'' 2024. [Online]. Available: \url{https://arxiv.org/abs/2307.15992}
\BIBentrySTDinterwordspacing

\bibitem{yoo2024advancingidentificationmultibitwatermark}
K.~Yoo, W.~Ahn, J.~Jang, and N.~Kwak, ``Robust multi-bit natural language
  watermarking through invariant features,'' in \emph{ACL}, 2023, pp.
  2092--2115.

\bibitem{kirchenbauer2024reliabilitywatermarkslargelanguage}
\BIBentryALTinterwordspacing
J.~Kirchenbauer, J.~Geiping, Y.~Wen, M.~Shu, K.~Saifullah, K.~Kong,
  K.~Fernando, A.~Saha, M.~Goldblum, and T.~Goldstein, ``On the reliability of
  watermarks for large language models,'' 2024. [Online]. Available:
  \url{https://arxiv.org/abs/2306.04634}
\BIBentrySTDinterwordspacing

\bibitem{zhao2023provablerobustwatermarkingaigenerated}
\BIBentryALTinterwordspacing
X.~Zhao, P.~Ananth, L.~Li, and Y.-X. Wang, ``Provable robust watermarking for
  ai-generated text,'' 2023. [Online]. Available:
  \url{https://arxiv.org/abs/2306.17439}
\BIBentrySTDinterwordspacing

\bibitem{hu2023unbiasedwatermarklargelanguage}
\BIBentryALTinterwordspacing
Z.~Hu, L.~Chen, X.~Wu, Y.~Wu, H.~Zhang, and H.~Huang, ``Unbiased watermark for
  large language models,'' 2023. [Online]. Available:
  \url{https://arxiv.org/abs/2310.10669}
\BIBentrySTDinterwordspacing

\bibitem{wu2024resilientaccessibledistributionpreservingwatermark}
\BIBentryALTinterwordspacing
Y.~Wu, Z.~Hu, J.~Guo, H.~Zhang, and H.~Huang, ``A resilient and accessible
  distribution-preserving watermark for large language models,'' 2024.
  [Online]. Available: \url{https://arxiv.org/abs/2310.07710}
\BIBentrySTDinterwordspacing

\bibitem{qu2025provablyrobustmultibitwatermarking}
\BIBentryALTinterwordspacing
W.~Qu, W.~Zheng, T.~Tao, D.~Yin, Y.~Jiang, Z.~Tian, W.~Zou, J.~Jia, and
  J.~Zhang, ``Provably robust multi-bit watermarking for ai-generated text,''
  2025. [Online]. Available: \url{https://arxiv.org/abs/2401.16820}
\BIBentrySTDinterwordspacing

\bibitem{fairoze2024publiclydetectablewatermarkinglanguagemodels}
\BIBentryALTinterwordspacing
J.~Fairoze, S.~Garg, S.~Jha, S.~Mahloujifar, M.~Mahmoody, and M.~Wang,
  ``Publicly-detectable watermarking for language models,'' 2024. [Online].
  Available: \url{https://arxiv.org/abs/2310.18491}
\BIBentrySTDinterwordspacing

\bibitem{liu2023unforgeable}
A.~Liu, L.~Pan, X.~Hu, S.~Li, L.~Wen, I.~King, and S.~Y. Philip, ``An
  unforgeable publicly verifiable watermark for large language models,'' in
  \emph{The Twelfth International Conference on Learning Representations},
  2023.

\bibitem{kothapalli2022nova}
A.~Kothapalli, S.~Setty, and I.~Tzialla, ``Nova: Recursive zero-knowledge
  arguments from folding schemes,'' in \emph{Annual International Cryptology
  Conference}.\hskip 1em plus 0.5em minus 0.4em\relax Springer, 2022, pp.
  359--388.

\bibitem{krishna2024paraphrasing}
K.~Krishna, Y.~Song, M.~Karpinska, J.~Wieting, and M.~Iyyer, ``Paraphrasing
  evades detectors of ai-generated text, but retrieval is an effective
  defense,'' \emph{Advances in Neural Information Processing Systems}, vol.~36,
  2024.

\bibitem{mitchell2023detectgpt}
E.~Mitchell, Y.~Lee, A.~Khazatsky, C.~D. Manning, and C.~Finn, ``Detectgpt:
  Zero-shot machine-generated text detection using probability curvature,'' in
  \emph{International Conference on Machine Learning}.\hskip 1em plus 0.5em
  minus 0.4em\relax PMLR, 2023, pp. 24\,950--24\,962.

\bibitem{verma-etal-2024-ghostbuster}
V.~Verma, E.~Fleisig, N.~Tomlin, and D.~Klein, ``Ghostbuster: Detecting text
  ghostwritten by large language models,'' in \emph{ACL}, Jun. 2024, pp.
  1702--1717.

\bibitem{hans2024spottingllmsbinocularszeroshot}
\BIBentryALTinterwordspacing
A.~Hans, A.~Schwarzschild, V.~Cherepanova, H.~Kazemi, A.~Saha, M.~Goldblum,
  J.~Geiping, and T.~Goldstein, ``Spotting llms with binoculars: Zero-shot
  detection of machine-generated text,'' 2024. [Online]. Available:
  \url{https://arxiv.org/abs/2401.12070}
\BIBentrySTDinterwordspacing

\bibitem{elkhatat2023evaluating}
A.~M. Elkhatat, K.~Elsaid, and S.~Almeer, ``Evaluating the efficacy of ai
  content detection tools in differentiating between human and ai-generated
  text,'' \emph{International Journal for Educational Integrity}, vol.~19,
  no.~1, p.~17, 2023.

\bibitem{zhao2024provable}
X.~Zhao, P.~V. Ananth, L.~Li, and Y.-X. Wang, ``Provable robust watermarking
  for {AI}-generated text,'' in \emph{The Twelfth International Conference on
  Learning Representations}, 2024.

\bibitem{pang2024freelunchllmwatermarking}
\BIBentryALTinterwordspacing
Q.~Pang, S.~Hu, W.~Zheng, and V.~Smith, ``No free lunch in llm watermarking:
  Trade-offs in watermarking design choices,'' 2024. [Online]. Available:
  \url{https://arxiv.org/abs/2402.16187}
\BIBentrySTDinterwordspacing

\bibitem{2811}
E.~Barker and J.~Kelsey, ``\BIBforeignlanguage{en}{Recommendation for random
  number generation using deterministic random bit generators},'' 2015-06-24
  2015.

\bibitem{tausworthe1965random}
R.~C. Tausworthe, ``Random numbers generated by linear recurrence modulo two,''
  \emph{Mathematics of Computation}, vol.~19, no.~90, pp. 201--209, 1965.

\bibitem{blum1986simple}
L.~Blum, M.~Blum, and M.~Shub, ``A simple unpredictable pseudo-random number
  generator,'' \emph{SIAM Journal on computing}, vol.~15, no.~2, pp. 364--383,
  1986.

\bibitem{matsumoto1998mersenne}
M.~Matsumoto and T.~Nishimura, ``Mersenne twister: a 623-dimensionally
  equidistributed uniform pseudo-random number generator,'' \emph{ACM
  Transactions on Modeling and Computer Simulation}, vol.~8, no.~1, pp. 3--30,
  1998.

\bibitem{marsaglia2003xorshift}
G.~Marsaglia, ``Xorshift rngs,'' \emph{Journal of Statistical software},
  vol.~8, pp. 1--6, 2003.

\bibitem{albrecht2016mimc}
M.~Albrecht, L.~Grassi, C.~Rechberger, A.~Roy, and T.~Tiessen, ``Mimc:
  Efficient encryption and cryptographic hashing with minimal multiplicative
  complexity,'' in \emph{International Conference on the Theory and Application
  of Cryptology and Information Security}.\hskip 1em plus 0.5em minus
  0.4em\relax Springer, 2016, pp. 191--219.

\bibitem{grassi2021poseidon}
L.~Grassi, D.~Khovratovich, C.~Rechberger, A.~Roy, and M.~Schofnegger,
  ``Poseidon: A new hash function for $\{$Zero-Knowledge$\}$ proof systems,''
  in \emph{30th USENIX Security Symposium}, 2021, pp. 519--535.

\bibitem{grassi2023poseidon2}
L.~Grassi, D.~Khovratovich, and M.~Schofnegger, ``Poseidon2: A faster version
  of the poseidon hash function,'' in \emph{International Conference on
  Cryptology in Africa}.\hskip 1em plus 0.5em minus 0.4em\relax Springer, 2023,
  pp. 177--203.

\bibitem{setty2012taking}
S.~Setty, V.~Vu, N.~Panpalia, B.~Braun, A.~J. Blumberg, and M.~Walfish,
  ``Taking proof-based verified computation a few steps closer to
  practicality,'' in \emph{21st USENIX Security Symposium}, 2012, pp. 253--268.

\bibitem{groth2016size}
J.~Groth, ``On the size of pairing-based non-interactive arguments,'' in
  \emph{Advances in Cryptology--EUROCRYPT}.\hskip 1em plus 0.5em minus
  0.4em\relax Springer, 2016, pp. 305--326.

\bibitem{PLONK}
\BIBentryALTinterwordspacing
A.~Gabizon, Z.~J. Williamson, and O.~Ciobotaru, ``{PLONK}: Permutations over
  lagrange-bases for oecumenical noninteractive arguments of knowledge,''
  Cryptology ePrint Archive, 2019. [Online]. Available:
  \url{https://eprint.iacr.org/2019/953}
\BIBentrySTDinterwordspacing

\bibitem{circom}
iden3, ``Circom,'' \url{https://github.com/iden3/circom}, 2021.

\bibitem{circomlib}
------, ``circomlib,'' \url{https://github.com/iden3/circomlib}, 2021.

\bibitem{halo2}
Zcash, ``halo2,'' \url{https://github.com/zcash/halo2}, 2022.

\bibitem{nova-scotia}
nalinbhardwaj, ``Nova-scotia,''
  \url{https://github.com/nalinbhardwaj/Nova-Scotia}, 2023.

\bibitem{nova}
microsoft, ``Nova,'' \url{https://github.com/microsoft/Nova}, 2022.

\bibitem{kim2016extended}
T.~Kim and R.~Barbulescu, ``Extended tower number field sieve: A new complexity
  for the medium prime case,'' in \emph{Annual international cryptology
  conference}.\hskip 1em plus 0.5em minus 0.4em\relax Springer, 2016, pp.
  543--571.

\bibitem{wood2014ethereum}
G.~Wood \emph{et~al.}, ``Ethereum: A secure decentralised generalised
  transaction ledger,'' \emph{Ethereum project yellow paper}, vol. 151, no.
  2014, pp. 1--32, 2014.

\bibitem{chen2017algorand}
\BIBentryALTinterwordspacing
J.~Chen and S.~Micali, ``Algorand,'' 2017. [Online]. Available:
  \url{https://arxiv.org/abs/1607.01341}
\BIBentrySTDinterwordspacing

\bibitem{webster1985design}
A.~F. Webster and S.~E. Tavares, ``On the design of s-boxes,'' in
  \emph{Conference on the theory and application of cryptographic
  techniques}.\hskip 1em plus 0.5em minus 0.4em\relax Springer, 1985, pp.
  523--534.

\bibitem{Avalanche_Alg}
\BIBentryALTinterwordspacing
A.~Sateesan. (2020) Analyze your hash functions: The avalanche metrics
  calculation. [Online]. Available:
  \url{https://arishs.medium.com/analyze-your-hash-functions-the-avalanche\\-metrics-calculation-767b7445ee6f}
\BIBentrySTDinterwordspacing

\bibitem{setty2020spartan}
S.~Setty, ``Spartan: Efficient and general-purpose zksnarks without trusted
  setup,'' in \emph{Annual International Cryptology Conference}.\hskip 1em plus
  0.5em minus 0.4em\relax Springer, 2020, pp. 704--737.

\end{thebibliography}

\end{document}